\documentclass[a4paper,fleqn,usenatbib,useAMS]{mnras}

\usepackage[dvipdfmx]{graphicx} %% to use \includegraphics
\usepackage[dvipdfmx]{color}
\usepackage{xspace}
\usepackage[dvipsnames]{xcolor} %% to use more colors (e.g. gray)

\newcommand{\sH}{$\Sigma_{\rm H}$\xspace}
\newcommand{\sstar}{$\Sigma_{\rm star}$\xspace}

\usepackage[T1]{fontenc}
\usepackage{ae,aecompl}

\usepackage{mathptmx}
\usepackage{txfonts}

\title[Gas and stellar spiral arms in M51]
{
Gas and stellar spiral arms and their offsets
in the grand-design spiral galaxy M51 
}
\author[Egusa et al.]
{Fumi Egusa$^{1,2}$\thanks{E-mail: fumi.egusa@nao.ac.jp},  
Erin Mentuch Cooper$^{3}$, 
Jin Koda$^{4}$, and Junichi Baba$^{5,6}$\\
$^{1}$Institute of Space and Astronautical Science, 
Japan Aerospace Exploration Agency, 
Sagamihara, Kanagawa 252-5210, Japan\\
$^{2}$Chile Observatory, National Astronomical Observatory of Japan, Mitaka, Tokyo 181-8588, Japan\\
$^{3}$Department of Astronomy, The University of Texas at Austin, Austin, TX 78712-1205, USA\\
$^{4}$Department of Physics and Astronomy, Stony Brook University, Stony Brook, NY 11794-3800, USA\\
$^{5}$Earth-Life Science Institute, Tokyo Institute of Technology, 
Meguro-ku, Tokyo 152-8550, Japan\\
$^{6}$Research Center for Space and Cosmic Evolution, Ehime University, Matsuyama, Ehime 790-8577, Japan
}

\date{Last updated ; in original form }

\pubyear{2016}

\begin{document}
\label{firstpage}
\pagerange{\pageref{firstpage}--\pageref{lastpage}}
\maketitle

%% NOTE: word limit = 250 words !!
\begin{abstract}
 Theoretical studies on the response of interstellar gas to 
a gravitational potential disc with a 
quasi-stationary
spiral arm pattern suggest 
that the gas experiences a sudden compression due to standing shock waves 
at
spiral arms.
 This mechanism, called a galactic shock wave, predicts that 
gas spiral arms move from downstream to upstream of stellar arms
with increasing radius inside a corotation radius.
 In order to investigate if this mechanism is at work
in the grand-design spiral galaxy M51,
we have measured azimuthal offsets between the peaks of stellar mass and 
gas mass distributions in its two spiral arms.
 The stellar mass distribution is created by the 
spatially resolved spectral energy distribution fitting 
to optical and near infrared images, while the gas mass distribution 
is obtained by high-resolution CO and HI data. 
 For the inner region ($r \leq 150''$), we find that one arm is consistent 
with the galactic shock while the other is not.
 For the outer region, results are less certain due to 
the narrower range of offset values, 
the weakness of stellar arms, and the smaller number of successful offset measurements.
 The results suggest that the nature of two inner spiral arms are different, 
which is likely due to an interaction with the companion galaxy.
\end{abstract}


\section{Introduction}\label{sec:intro}
 The nature of stellar and gaseous spiral arms in galactic discs 
has been studied about 50 years and is still being discussed actively.
 In order to avoid the winding problem, in which differentially rotating spiral arms quickly
become too tightly wound to hold the spiral structure, 
\citet{LS64} proposed a density wave hypothesis, 
in which the spiral pattern is a density wave propagating the disc 
and is a quasi-stationary structure 
for at least several rotations, i.e.\ $\sim 1$ Gyr.
%% v_rot=200km/s at r=8kpc -> one rotaion is ~240 Myr.
 Based on this hypothesis,
\citet{Fuji68} and \citet{RobW69} calculated a non-linear response of 
interstellar gas in a gravitational potential 
with a fixed spiral arm structure.
 Assuming isothermal gas with tightly wound spiral arms, 
they found a shock front appears upstream of the potential minima 
if the spiral arm perturbation is strong enough.
 Their results agree well with the observational fact 
that typical width of gas or dust in spiral arms is usually much narrower 
than that of the stellar mass seen in near-infrared images.
 Furthermore, a sharp increase of gas density at the shock front 
is recognized as a trigger of star formation in spiral arms.
 This picture,
i.e., the gas response under a fixed or quasi-stationary spiral potential, 
is called galactic shock wave theory.
%% Wood75: gas without self gravity, with tightly wound stellar spiral
 In the case of tightly wound stellar spiral arms, 
\citet{Wood75} presented that the galactic shocks appear within a few 100 Myr.
 This result suggests that the galactic shock needs a spiral potential 
to be stationary for more than this time-scale.

 Since these pioneering works, many theoretical studies have been done 
to calculate the gas response for a fixed potential with different conditions and assumptions.
%% Shu72:
%% Shu73: 
%% Ishi84: 2D numerical calculations, i=7deg with energy gain&loss,
%%  magnetic field, self gravity ignored. 
%%  density contrast smaller for adiabatic shock than isothermal
%%  gas density peak may not coincide with the shock front
%% Lub86: 1D, WKB, gas self gravity included for the first time, at Solar radius
%%  stellar peak at gas peak (downstream), when gas density is high (potential not fixed)
%%  shock features disappear when gas fraction is high $\gtrsim 8$\%
%% KO02: 1D
%% Wada08: fixed spiral but shocks are unstable?
%% GC04: isothermal gas for tightly wound spirals, offset (gas-star) is a decreasing function of radius and becomes -90deg (Theta=-PI) at CR (F17,18,20), only r<R_CR is presented
%% Mar09b: semianalytic (sin(i)=0.1?) and MHD (i=15.5deg) calculations, shock moves toward upstream with increasing radius in both cases (shock more downstream in MHD)
%% KimY14: isothermal, F=5,10,20%, i=20, w/ & w/o self gravity, difference of pitch angles between stellar gans gas arms depends on the pattern speed
%% LeeWK14: 1D, F=8%, i=21 (ref model), MHD, stronger self gravity (or weaker B-field?) -> shock at downstream
%% Baba15: i=10, 20; F=2, 5%; isothermal or multiphase (not "tightly wound")
 In this paper, we call such models as steady spiral models.
 In the case of isothermal gas with tightly wound spiral arms, 
\citet{GC04} presented that the shock front should move 
monotonically from downstream to upstream with increasing radius. 
 This trend holds inside the corotation resonance, 
i.e.\ where gas and stars rotate faster than the spiral pattern.
 A similar trend is found in other cases with more open spiral arms \citep{KimY14,Baba15} 
and with magnetic field \citep{Mar09b}.
 This trend is likely 
because that a relative velocity between gas and a spiral pattern depends on radius, 
and thus should be an universal and model-independent characteristic of the galactic shock wave.
 The effect of gas self gravity to the shock location has also been investigated. 
 Some studies based on one dimensional (1D) calculations suggested that 
a higher gas fraction, i.e.\ a stronger self gravity, 
pushes the gas peak downstream \citep{Lub86,KO02,LeeWK14}.
 Furthermore, these studies claimed that the galactic shock is not formed 
when the gas density is too high.
 Meanwhile, recent two dimensional (2D) calculations by \citet{Baba15} presented that 
gas peak positions with and without self gravity do not significantly differ when the potential is fixed. 
 This might be because that models are different (e.g.\ 1D vs 2D), 
shift due to self gravity is too small, 
and/or stellar feedback included in Baba's model counteracts with self gravity.
 Therefore, how gas self gravity changes the gas peak positions is not clear yet.

%% simulations of stellar disk suggest transient nature of spiral arms (without bar or tidal)?
%% Wada11: 3D N-body + SPH with self gravity, heating & cooling, star formation & feedback
%%  for isolated open spiral
%% Dob10: 3D SPH simulation for M51, fiducial: 1% gas of disk with T=10^4K (isothermal)
%%  no offsets between stars and gas
%% Dob10b: age distribution of notional star clusters in 4 types of galaxies
%%  clear dependence seen only in tightly wound density waves and bar
%% Stru11: long-lived spiral arms with interaction, isothermal 10^4K gas, 
%%  no systematic offsets between stars and gas
%% Baba13: 3D N-body for stellar arms without bar & tidal, analyzed stellar orbits 
%% Pett16: gas slightly ($\sim 10^\circ$) upstream of stars (Fig. 6)
 On the other hand, recent numerical simulations 
for dynamic stellar discs have suggested that
spiral structures in isolated galaxies are multi-armed and not stationary 
\citep[e.g.][]{Fujii11,Baba13}.
 In this paper, we call such models as dynamic spiral models.
 In these cases, spiral patterns almost corotate with the materials, 
and thus a relative velocity of gas to the pattern is small compared 
to the case of steady spiral models.
 As a result, no systematic offsets are seen in these simulations 
\citep{Dob08a,Wada11,Baba15}.
 For interacting galaxies, \citet{Dob10} and \citet{Stru11} studied 
a time evolution of disc structures.
 While the lifetime of spiral arms was longer ($\sim$ a few 100 Myr)
compared to isolated galaxies, 
no systematic offsets were found between gas and stellar arms.
 \citet{Pett16} also investigated tidally interacting galaxies 
with a wider range of parameters, and found that gas arms appear 
slightly upstream of stellar arms for a few 100 Myr, 
when the spiral arm structure is enhanced due to the interaction.
 For more details of theoretical calculations on spiral structures, 
see a recent review by \citet{DB14}.

 As described above, theoretical studies suggested that 
the radial trends of gas-star offsets differ between the two spiral models. 
 To clarify the difference, here we summarize the results of 
\citet{Baba15} who investigated the location of gas and stellar spiral arms, 
by performing high resolution simulations of
steady and dynamic spiral models.
 For steady spiral models, stellar spiral arms were assumed to be 
a rigidly-rotating potential.
 They ran four hydrodynamic simulations with parameters in combination of 
two pitch angles ($10^\circ$ and $20^\circ$), 
two spiral arm strengths (2\% and 5\%), and with or without gas self gravity.
 Similarly to the  
aforementioned steady spiral models,
gas spiral arms appear downstream of stellar spiral arms at inner radii 
and move to upstream at outer radii.
 While absolute offset values vary with adopted parameters, 
this radial trend itself does not change.
 For dynamic spiral models, 
they simulated a MW-like galaxy with 
$N$-body/hydrodynamic calculations including gas self gravity.
 During the course of the simulation, a bar in the centre developed.
 On average, spiral arms are more open ($\sim 25^\circ$--$30^\circ$) and 
stronger ($\sim 5$\%--8\%) than in the steady spiral cases.
 Similarly to the other dynamic spiral models, 
gas spiral arms appear very close to the stellar spiral arms.
 No systematic offsets between the gas and stellar arms are found, 
and the situation does not depend on whether there is the bar or not.
 From these results for steady and dynamic spiral models, 
the authors claimed that the steadiness of spiral arms is the key 
to the offset dependence on radius.
 In other words, the radial dependence of gas-star offsets 
is suggested to be a useful observational tool to distinguish 
steady and dynamic spiral arms.

 Based on this theoretical suggestion, 
we measure the gas-star offsets from the location of 
gas density peaks relative to stellar density peaks, 
both of which are estimated from observational data sets.
 The goal of this study is to investigate if gas spiral arms shift 
from downstream to upstream with increasing radius, 
as predicted by the steady spiral and galactic shock models.

 A prototypical grand-design spiral galaxy, M51 (or NGC 5194), 
is selected as a target because the arm to interarm contrast 
has been known to be large at a wide range of wavelength.
 The large contrast in stellar potential is necessary to form shock waves 
while that in gas density indicates the shock likely exists.
 This galaxy is also known as the whirlpool galaxy and 
is interacting with a companion galaxy, M51b or NGC 5195.
 We adopt parameters of M51 used in \citet{Egu13} 
and list them in Table \ref{tb:m51}.
 Their notation of two spiral arms, arm1 being to extend to the opposite side of M51b 
and arm2 being connected to M51b, is also adopted.

\begin{table}
	\caption{Adopted parameters of M51}\label{tb:m51}
	\begin{tabular}{lc}
		\hline
		R.A. (J2000) & $13:29:52.711$\\
		Decl. (J2000) & $+47:11:42.62$\\
		P.A. & $169^\circ$\\
		Incl. & $24^\circ$\\
		Distance & 8.4 Mpc\\
		\hline
	\end{tabular}
\end{table}

\section{Data \& analysis}\label{sec:data}
\subsection{Total H gas density}
 For atomic H gas, the integrated intensity map of the HI 21cm emission has been 
obtained from \citet{THINGS},
whose angular resolution 
(full width at half maximum (FWHM) of a gaussian beam)
is $5.8''\times 5.6''$.  
 For molecular H$_2$ gas, the integrated intensity map of the $^{12}$CO(1--0) emission
has been obtained from \citet{Koda09}, 
whose angular resolution is $3.7''\times 2.8''$. 
 The factor $X_{\rm CO}=2\times 10^{20}$ [H$_2$/cm$^2$/(K km/s)] 
has been adopted to convert the CO integrated intensity into the H$_2$ gas surface density.
 Discussion about a variation of this factor is given in Appendix \ref{sec:Xco}.

 Both maps have been 
convolved with 2D gaussian profiles and then regridded
so that the angular resolution become $6''$ ($\sim 240$ pc at the distance of M51) 
and the pixel size become $2''$.
 We have calculated the total surface density of H, \sH, 
by adding these two maps using the conversion factor.
 As already presented by \citet{Koda09}, 
\sH is dominated by the H$_2$ gas for most of the M51 disc.
 The ionized gas and warm H$_2$ gas components are neglected in this study.

\subsection{Stellar mass density}
\begin{figure*}
\includegraphics[width=0.48\linewidth]{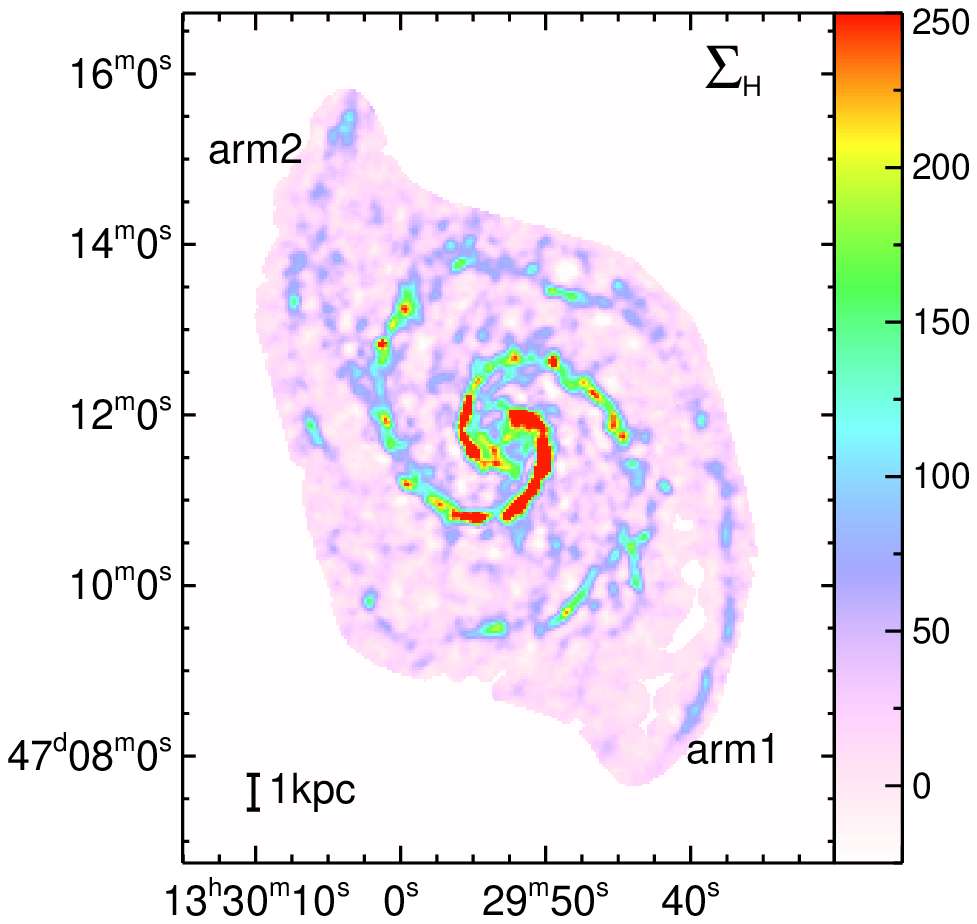}
\includegraphics[width=0.48\linewidth]{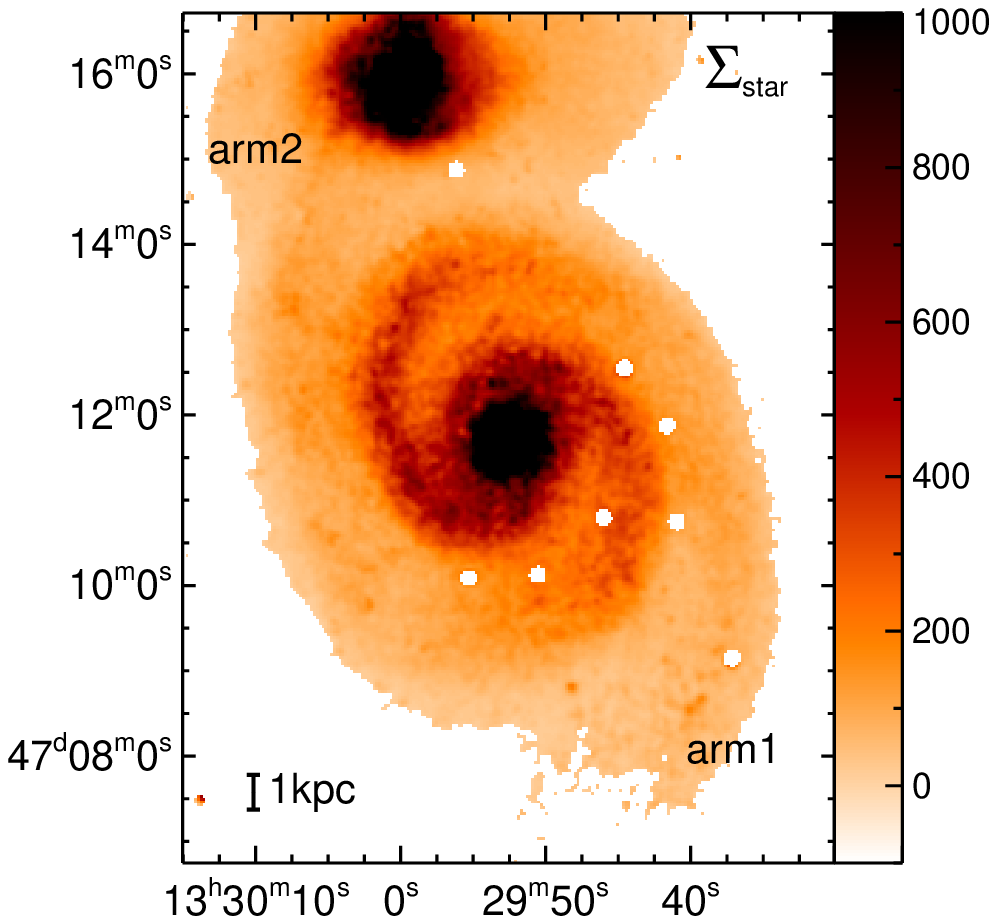}
\caption{Smoothed and regridded images of \sH (left) and \sstar (right) of M51.
The coordinates are right ascension and declination (J2000), 
and the unit for colour scale is [$M_\odot$/pc$^2$] for both images.
}
\label{fig:showfits}
\end{figure*}

 \citet{MenC12} created a spatially resolved map of the stellar mass distribution in M51 
by fitting spectral energy distribution (SED) models to optical through 
near-infrared data.
 While the original angular resolution of the data presented in their paper was 
$28''$ to match the lower resolution of 
infrared data, here we use a modified analysis 
at higher spatial resolution of $4''$ and limit our interest to the stellar mass distribution within the galaxy. 
 Plausible MW stars have been identified and masked out.
 The map of stellar mass surface density, \sstar, 
has also been smoothed and regridded to match the \sH map.
 These images are presented in Figure \ref{fig:showfits}.

 While the \sstar map resembles near-infrared (e.g.\ $K$-band) images, 
it is more reliable as the near infrared luminosity can still be affected by young stars 
and interstellar dust features, 
and thus the mass-to-luminosity ratio is not constant.
 Effect of attenuation to \sstar is discussed in Appendix \ref{sec:Av}.

\subsection{Phase diagrams}
\begin{figure*}
\includegraphics[width=\linewidth]{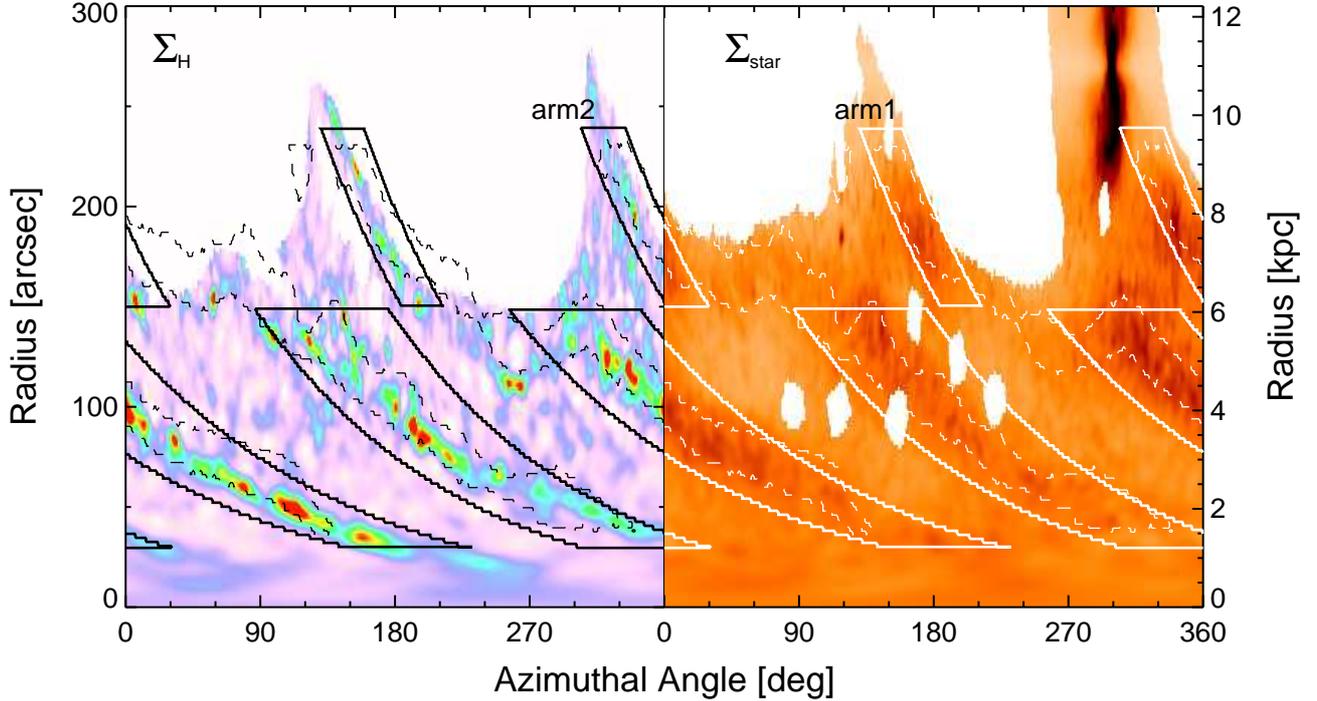}
\caption{Phase diagrams of \sH (left) and \sstar (right) of M51.
Each image is normalized by its azimuthal average at each radius 
in order to emphasize the spiral arm structure.
Thin dashed contours indicate the arm definition by \citet{Egu13}, 
while thick solid contours indicate new definition adopted in this paper 
(see \S \ref{sec:armdef}).
}
\label{fig:phase}
\end{figure*}

 The matched images have been deprojected to a face-on view 
and then transformed into the polar grid, i.e.\ (radius, azimuthal angle) 
$= (r,\theta)$, coordinates.
 Parameters listed in Table \ref{tb:m51} have been used in this transformation.
 We show these phase diagrams in Figure \ref{fig:phase}.
 The azimuthal angle starts at the eastern side of the minor axis (i.e.\ P.A. $= 79^\circ$)
and increases counter-clock wise, which is the same direction of the material flow 
assuming trailing arms.
 As a consequence, a larger azimuthal angle corresponds to the downstream side 
inside the corotation resonance.

\subsection{Arm definition}\label{sec:armdef}
 In order to avoid picking up peaks in interarm regions, 
we define arm regions based on the position of stellar arms.
 As an initial step, the arm definition from \citet{Egu13}, 
which is based on a stellar brightness map from \citet{MenC12}, is adopted.
 This definition is presented by the thin dashed contours in Figure \ref{fig:phase}.
 We regrid the \sstar phase diagram so that the radial interval becomes $4''$ 
and create azimuthal profile at each radius.
 For each arm, the peak of the profile within the defined arm region is identified at each radius.

 We then fit a logarithmic spiral 
$$ \theta = \frac{-1}{\tan{(i_{\rm pitch})}} \ln{r} + \theta_0$$ 
to the peak positions, assuming the pitch angle $i_{\rm pitch}$ is constant.
 Since both of the two arms in M51 show a break/kink at $r\sim 150''$, 
inner ($r\leq 150''$) and outer ($r>150''$) arms are fitted separately, 
i.e.\ four individual fittings are performed.
 We define arm regions as within the azimuthal range of 
$\pm 45^\circ$ (inner) or $\pm 15^\circ$ (outer)
from the stellar logarithmic spiral arms.
 These azimuthal ranges are determined so that they include major peak positions 
seen in the \sstar and \sH phase diagrams.
 This new arm definition is indicated by thick solid contours in Figure \ref{fig:phase}.
 The arm1 starts and ends at an azimuthal angle of $\sim 180^\circ$, 
while arm2 starts and ends at $\sim 330^\circ$.
 Dependence of results on this arm definition is discussed in Appendix \ref{sec:params}.

\subsection{Locating peaks}\label{sec:ploc}
 Here we describe how we define the location of \sH and \sstar peaks 
from their azimuthal profiles using two different methods.
 As for \sstar, \sH azimuthal profiles are created at $4''$ radial interval.

\subsubsection{Finding peaks}\label{sec:peakfind}
 Similarly to \S \ref{sec:armdef}, 
we locate peaks of azimuthal profiles 
within the new arm regions at each radial bin.
 We confirm that stellar 
peak positions are consistent between 
the initial and new arm definitions except for the inner- and outer-most regions 
and around the break.
 By definition, this method is very sensitive to local peaks.
 One caveat is that it could pick up false peaks due to noises.

 The logarithmic spiral fitting is performed again to these \sstar and \sH 
peak positions.
 The best-fitting pitch angles are listed in Table \ref{tb:pitch}.
 Our results for inner gas arms are consistent with \citet{Miya14}, 
who derived $i_{\rm pitch} = 19^\circ \pm 1^\circ$ for $r=40''$--$140''$ 
from the $^{12}$CO(1--0) map of M51 obtained with the NRO45m telescope.

%% fitting log-spiral to peak positions in _new.dat
\begin{table}
	\caption{Pitch angles by fitting logarithmic spiral to gas and stellar arms [degree]}\label{tb:pitch}
	\begin{tabular}{lcc}
		\hline
		\multicolumn{3}{c}{inner arms: $r=30''$--$150''$ }\\
		\hline
		& gas & star \\
		arm1 & $19.9\pm 0.3$ & $19.3\pm 0.5$ \\
		arm2 & $18.8\pm 0.6$ & $23.4\pm 0.9$ \\
		\hline
		\multicolumn{3}{c}{outer arms: $r=151''$--$220''$ }\\
		\hline
		& gas & star \\
		arm1 & $27.8_{-1.2}^{+1.3}$ & $37.4_{-3.0}^{+3.5}$ \\
		arm2 & $27.6_{-1.4}^{+1.5}$ & $32.4_{-2.4}^{+2.8}$ \\
		\hline
	\end{tabular}
\end{table}

\subsubsection{Fitting Gaussian profiles}\label{sec:gfit}
 \citet{Dob10} fit two-gaussian profiles to 
azimuthal profiles of simulated gas and stellar disc
in order to locate two spiral arms at a radial interval of 1 kpc.
 We basically follow their procedure except that we use 
the radial interval of $4''$. 
 If fitted peaks fall outside the defined arm regions, we exclude them from the following analysis.
 Widths of stellar and gas spiral arms from the fitting are
presented in \S \ref{sec:stellararm} and \ref{sec:gasarm}, respectively, 
and are used to estimate 
uncertainties in offset measurements (\S \ref{sec:err}).
 Compared to the peak-finding method, it is more sensitive to 
broader profiles.
 One caveat is that this fitting may not be appropriate 
when azimuthal profiles are not symmetric.

\subsubsection{Comparison}
\begin{figure}
\includegraphics[width=\linewidth]{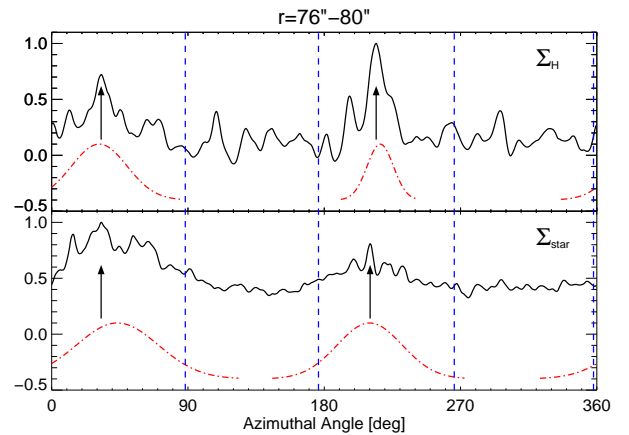}
\caption{Azimuthal profiles of \sH (top) and \sstar (bottom) at a radial bin of 
$r=76''$--$80''$.
The vertical axis is in arbitrary unit.
Arrows indicate the position of identified peaks 
and blue dashed vertical lines indicate the boundary of arm regions at this radius.
Fitted gaussian profiles are presented by red dashed-dotted lines.
}
\label{fig:showpeak}
\end{figure}

\begin{figure*}
\includegraphics[width=\linewidth]{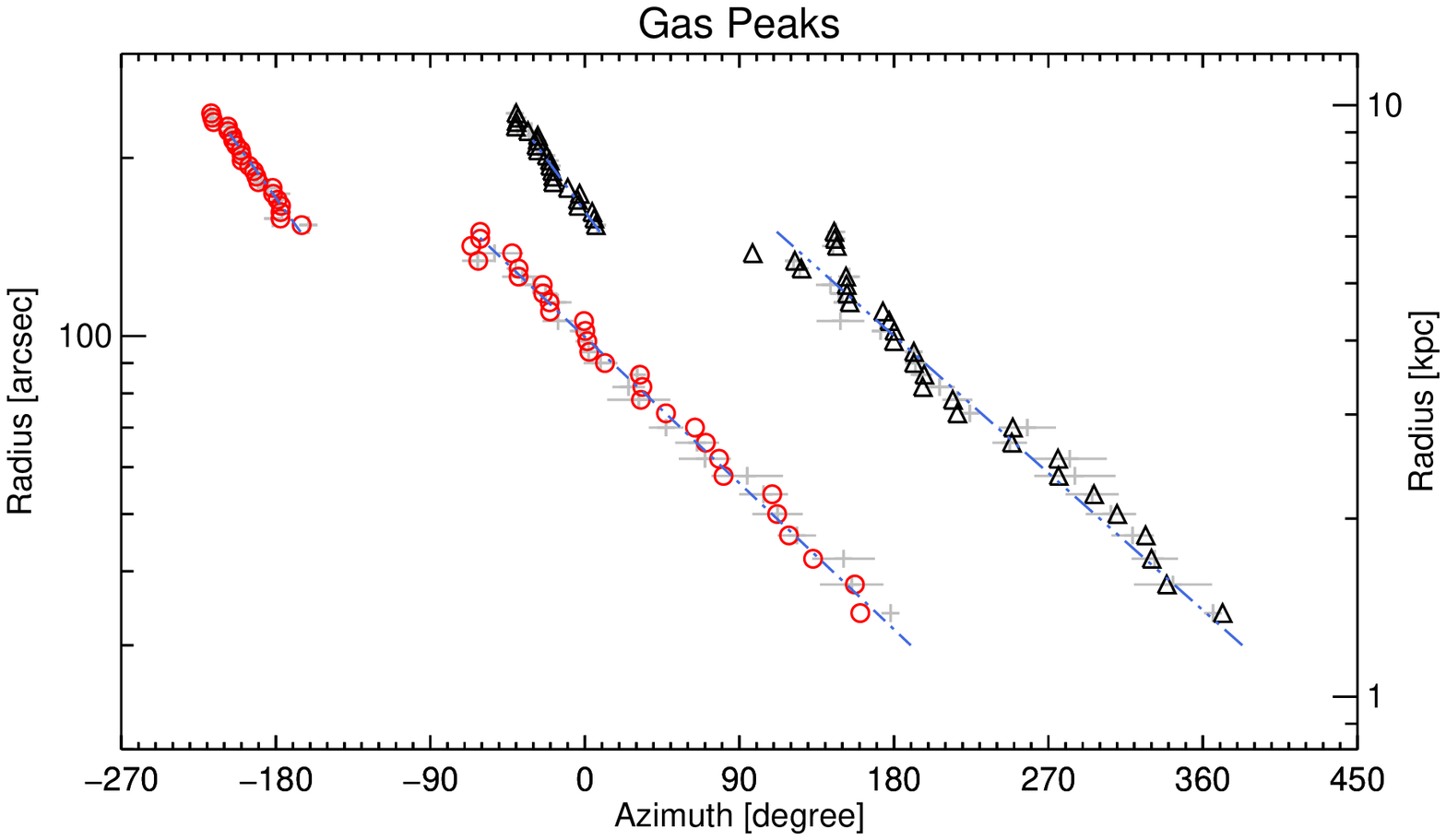}
\includegraphics[width=\linewidth]{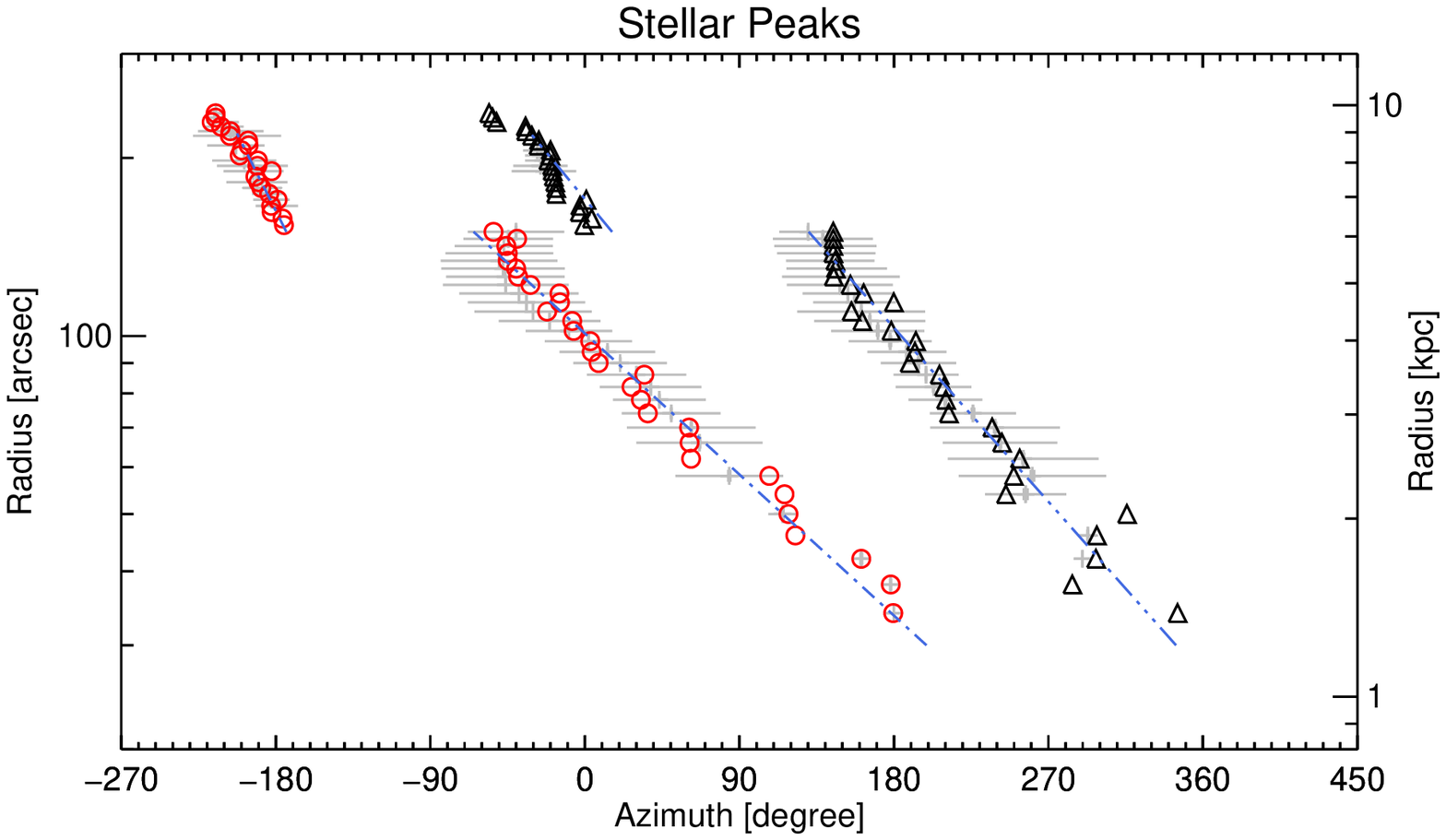}
\caption{Location of gas (top) and stellar (bottom)
spiral arms by different methods.
Red circles and black triangles are peaks of azimuthal profiles
within the arm regions.
Blue dot-dashed lines are logarithmic spiral arms
fitted to these peaks.
Gray crosses indicate peak positions from the gaussian fit. 
Thin horizontal bars indicate the width ($\pm \sigma$) of a fitted gaussian profile. 
Thick bars indicate the uncertainty of the peak but they are generally too small to be visible 
in this plot.
Gauss-fit peaks outside the arm regions are excluded.
}
\label{fig:peaks}
\end{figure*}

 An example of azimuthal profiles at a radial bin of 
$r=76''$--$80''$ is shown in Figure \ref{fig:showpeak}.
 The arm regions at this radius are indicated by the blue vertical dashed lines.
 The solid arrows indicate identified peaks within the arm regions, 
while the red dashed-dotted curves present fitted gaussian profiles. 
 Overall comparison with 
the peak-finding method and gauss-fit method
is presented in Figure \ref{fig:peaks}, together with the log-spiral fit.
 Note that in this figure the spiral arms are unwrapped and thus 
the azimuthal range is from $-270^\circ$ to 
$450^\circ$.
 In addition, radius is in logarithmic scale.
 The fitted logarithmic spirals are indicated by blue dot(s)-dashed lines.
 The identified
peaks are plotted by open circles and triangles for arm1 and arm2, respectively.
 The results of gaussian fitting is presented by gray horizontal crosses.
 The length of thin horizontal bars correspond to the width ($\pm 1\sigma$) of 
fitted gaussian profiles.
 Thick bars indicate the uncertainty of the peak position but 
they are generally too small to be visible in this plot.
 From these two figures, we find that the above two methods give generally consistent arm locations.

\subsection{Measuring offsets}\label{sec:measureofs}
 We define offsets as azimuth(H gas)$-$azimuth(star), 
so that the gas shock 
at a leading side of the stellar peak should result in a negative offset.
 Inside the corotation resonance, 
the leading side corresponds to the upstream side.

 For each of the peak locating methods, 
we measure offsets from all peak positions.
 In addition, we identify ``matched peaks'' where 
difference between the peaks from the two methods 
is smaller than the $\sigma$ of the fitted gaussian profile, 
i.e.\ the arm width.

 Uncertainties in measured offsets are estimated in \S \ref{sec:err}.

\section{Results}
 In this section, we first present an overview of stellar and gas spiral arms 
in \S \ref{sec:stellararm} and \ref{sec:gasarm}, respectively, focusing on 
the signal-to-noise (S/N) ratio of identified peaks 
and on the spiral arm width.
 The former is estimated by 
the ratio of peak brightness to RMS in interarm regions at each radius,
while the latter is measured from the gaussian fitting 
(\S \ref{sec:gfit}).
 For gas spiral arms, their locations are compared with a prediction 
by the galactic shock models.
 In \S \ref{sec:ofs}, the gas-star offsets  
and their radial dependence
are presented.
 We also discuss how results depend on the two peak locating methods.
 Offset dependence on the self-gravity strength 
is given in Appendix \ref{sec:alpha}.

%% Gar93a: arm/interarm from CO(2-1) increases with radius up to r~140"? peak at r~50", dip at r~90", and then increase??
%% NIR contrast increases w/ radius up to r~140''

\subsection{Stellar spiral arms}\label{sec:stellararm}
\begin{figure*}
\includegraphics[trim=0 0 0 0,clip,width=0.475\linewidth]{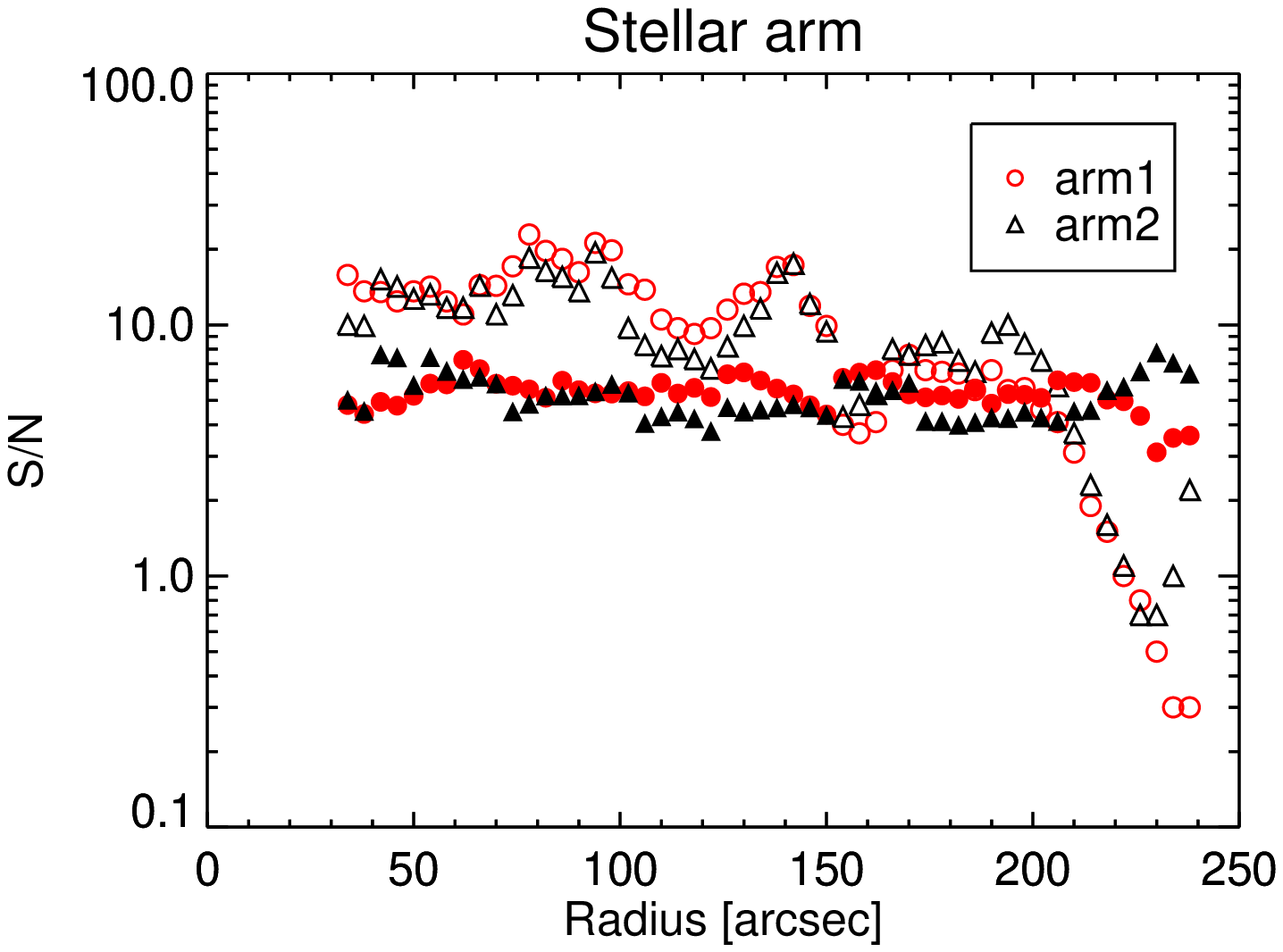}
\includegraphics[trim=0 0 0 0,clip,width=0.475\linewidth]{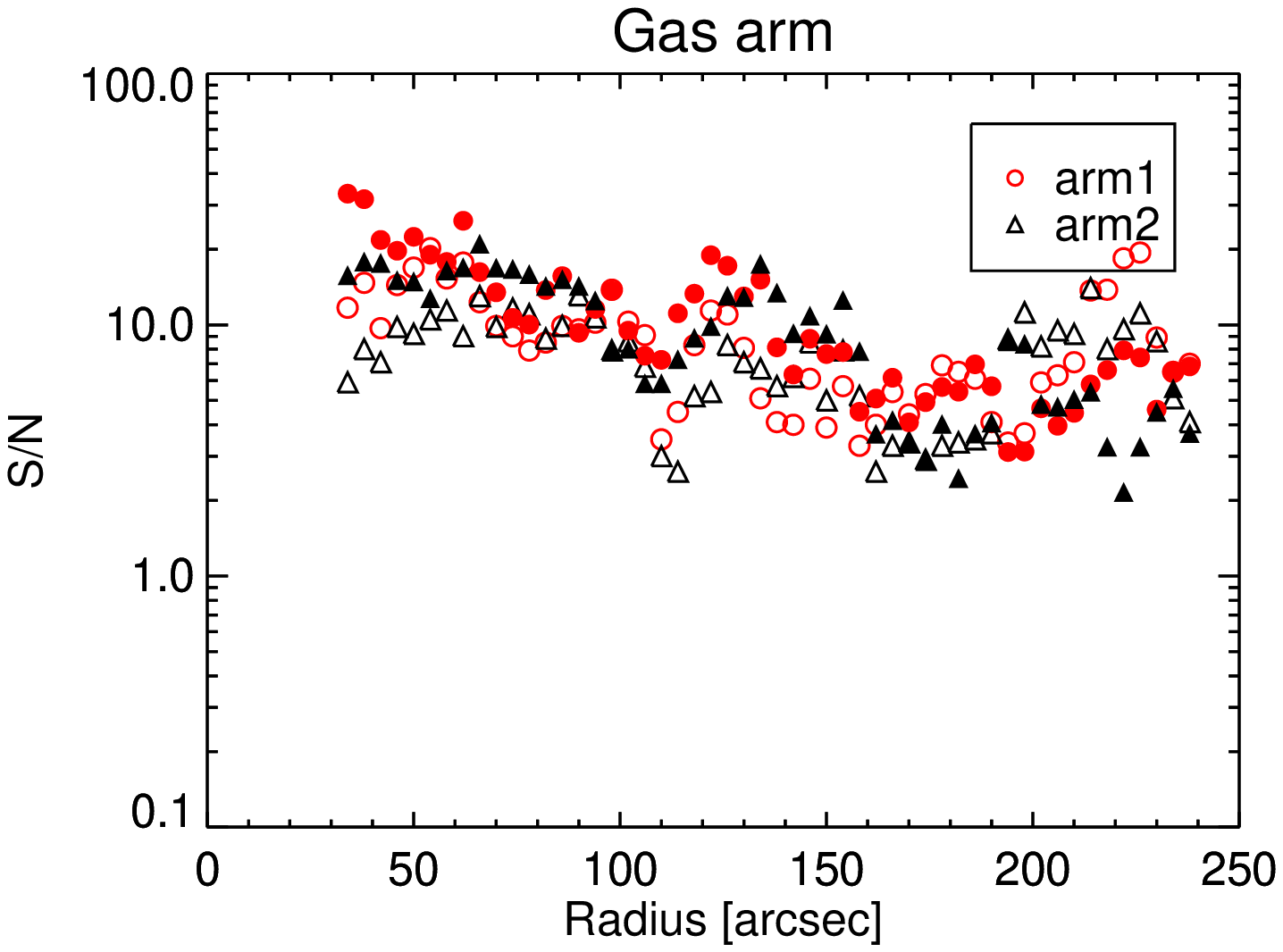}
\caption{
Signal-to-noise ratio (S/N) for stellar (left) and gas (right) spiral arms.
Open symbols indicate S/N values estimated from the ratio of peak brightness 
to the RMS of interarm regions.
Filled symbols indicate S/N values estimated from the stellar mass models at lower resolution 
for stellar arms and those estimated from the CO integrated intensity map 
and its error map for gas arms.
}
\label{fig:plot_sn}
\end{figure*}

 The S/N is estimated to be $\sim 10$ for inner arms, while is $\sim 5$ for outer arms 
(Figure \ref{fig:plot_sn}, left, open symbols).
 We find S/N becomes $< 2$ for $r>220''$, which is due to 
smaller interarm area with available data and to this area being affected by 
the brightness of the companion galaxy.
 This outermost region is thus excluded from the following analysis.
 In addition, we estimate the S/N at peak positions 
from the stellar mass surface density and its uncertainty at lower ($28''$) resolution 
derived by \citet{MenC12}, because the uncertainty map at $4''$ resolution is not available.
 These S/N values are presented by the filled symbols in the left panel of Figure \ref{fig:plot_sn}, 
and are almost constant (${\rm S/N} \simeq 5$) across the current radial range.

 Typical arm widths in azimuthal angle are $\simeq 60$ degree (FWHM) for inner arms 
and $\simeq 30$ degree for outer arms. 
 They are generally larger than those of gas spiral arms (Figure \ref{fig:peaks}, 
thin gray horizontal bars).

\subsection{Gas spiral arms}\label{sec:gasarm}
 The S/N is estimated to be between 3 and 20 (Figure \ref{fig:plot_sn}, right, open symbols), 
and has a weak trend of decreasing with radius.
 While gas arms appear more prominent than stellar arms, 
their S/N values do not differ significantly.
 This is probably due to the interarm structures (e.g.\ feathers) 
only remarkable at \sH map.
 In addition, we estimate the S/N at peak positions 
from the CO integrated intensity map and its uncertainty, 
because the uncertainty map for HI is not available.
 These S/N values are presented by the filled symbols in the right panel of Figure \ref{fig:plot_sn}, 
and are consistent with those estimated from the peak/rms ratio.

 Typical arm widths in azimuthal angle are $\simeq 30$ degree (FWHM) for inner arms 
and $\simeq 5$ degree for outer arms. 
 They are generally smaller than those of stellar spiral arms (Figure \ref{fig:peaks}, 
thin gray horizontal bars).

\textcolor{black}{
\subsubsection{Comparison with model-predicted locations}\label{sec:gasloc}
}
 In the case of galactic shock waves, 
gas spiral arms are expected to move from downstream to upstream 
with increasing radius inside the corotation.
 While this radial trend itself does not change, 
absolute offset values between the gas and stellar arms 
depend on models and parameters.
 Azimuthal offsets predicted by \citet{GC04} range from $-30$ to 0 degree 
and monotonically decrease with radius. 
%% LeeWK14: only at 1 radius, no discussion on radial dependence
 Here we take the result of \citet{Baba15} for comparison
and adopt the strong spiral arm (``F05'' in their model names) cases, 
since they are more suitable for M51 than the weak spiral case.
 The panels (B)--(D) of their Figure 2 present that
gas peak positions do not significantly differ between the models.
 The offset appears to linearly depend on radius and 
moves from $-45^\circ$ to $+45^\circ$ 
with increasing radius from 1 kpc to 9 kpc.
 Here we assume that this dependence should scale with 
the corotation radius ($R_{\rm CR}$), which is set to be 10 kpc in the models.
 Given that their the definition of offset direction is opposite to that in this paper, 
we adopt offset(gas-star) $=-90/0.8(r/R_{\rm CR}-0.5)$ degree 
within the range of $r=0.1R_{\rm CR}$--$0.9R_{\rm CR}$, 
as a prediction of the galactic shock waves.
%% gas spiral arm look logarithmic in \citet{GC04}, 
%% while close to linear in \citet{KimY14}. (but they measure pitch angle)

 For M51, there are many estimates of the corotation radius, 
most of which can be categorized into three:\footnote{
 A recent study on gas dynamics suggests 
the corotation at $r=100''$ \citep{Que16}.
 We do not include this possibility as galactic shock models 
outside the corotation are not available.
}
(i) $2.1'$--$2.2'$ \citep{Tul74c,Elm92,Vog93}, 
(ii) $2.7'$--$2.9'$ \citep{Gar93b,Egu09}, and
(iii) $5'$ \citep{Kuno95,Oey03}.
 The former two correspond to just inside and outside the break
of spiral arms, while the latter corresponds to the location of the companion galaxy.
 Furthermore, \citet{Meidt08b} suggested that 
the pattern speed in M51 may decrease with radius 
and that no corotation appears within the disk.
 We here test three constant $R_{\rm CR}$ cases, taking into account 
the area only inside each corotation radius.

\begin{figure*}
\includegraphics[trim=0 0 0 20,clip,width=0.32\linewidth]{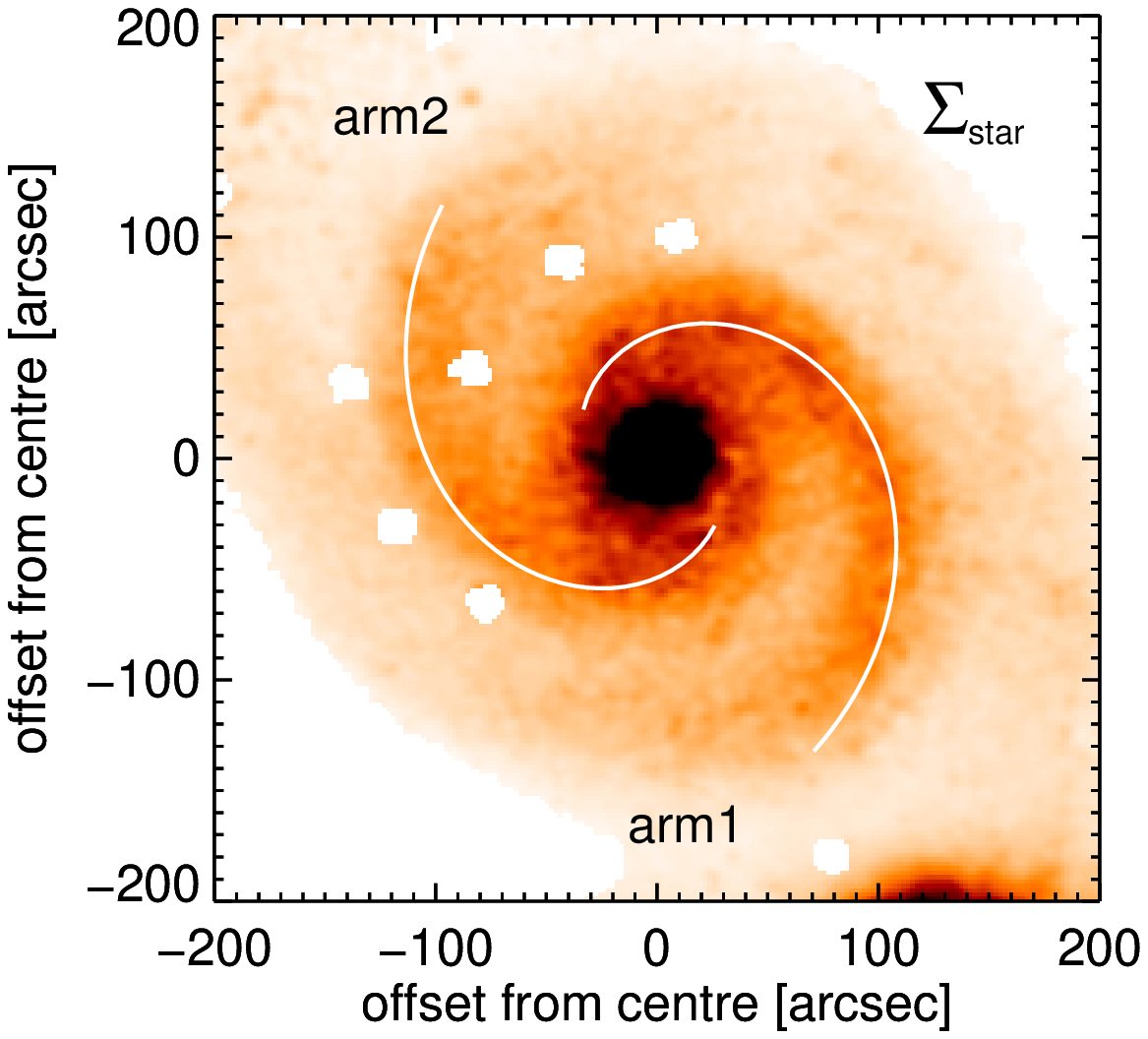}
\includegraphics[trim=0 0 0 20,clip,width=0.32\linewidth]{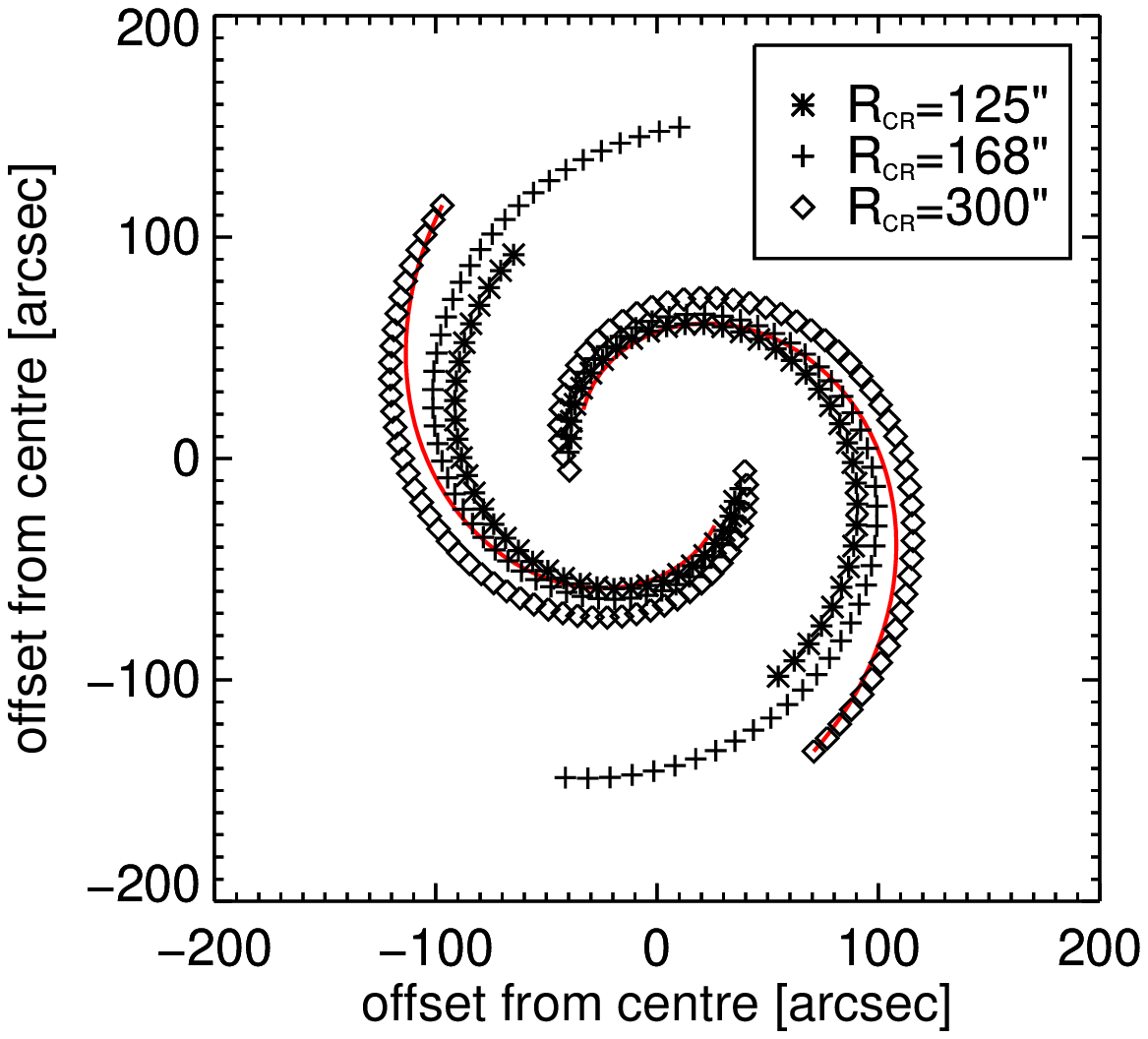}
\includegraphics[trim=0 0 0 20,clip,width=0.32\linewidth]{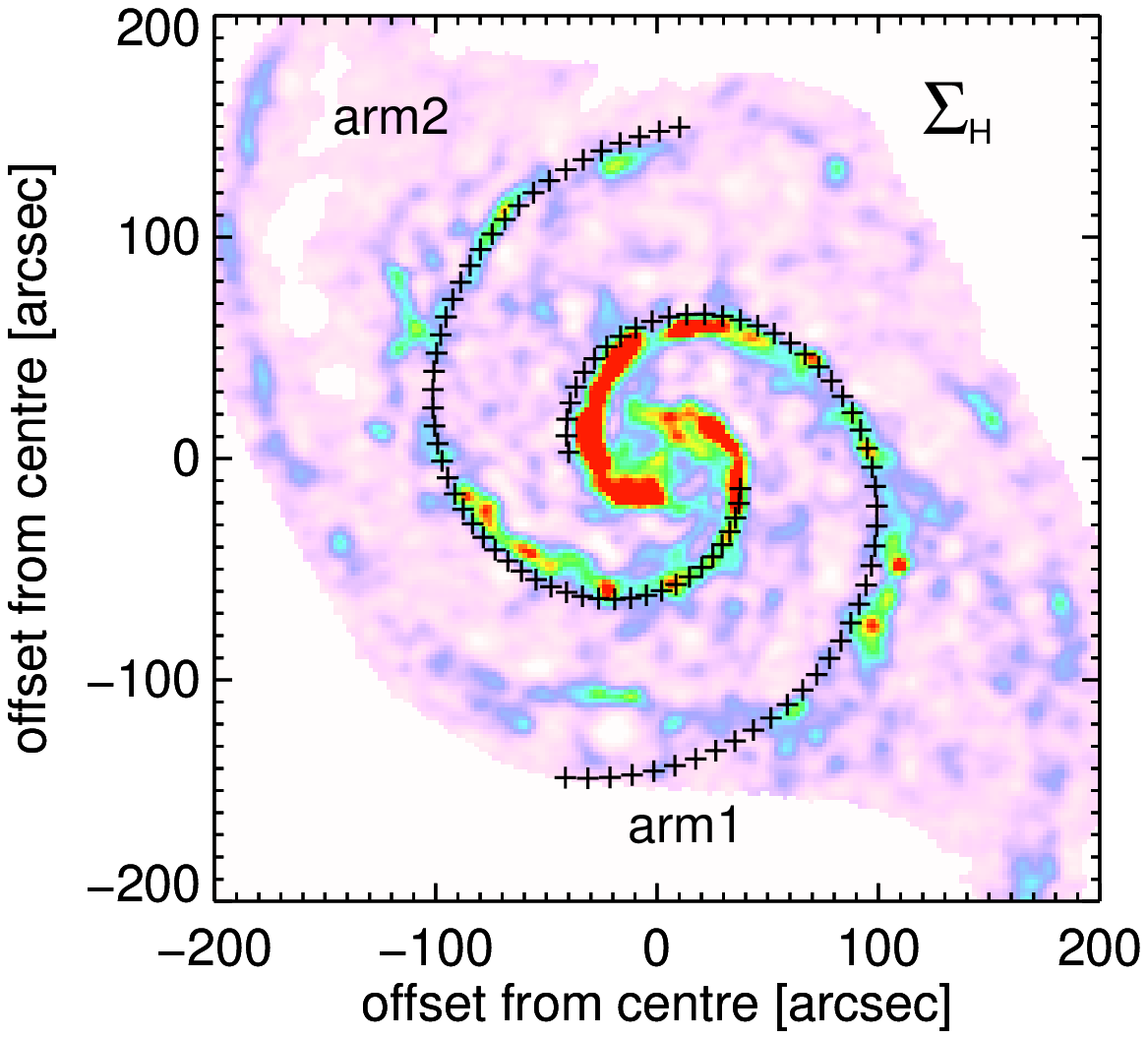}
\caption{
Left: fitted log-spiral stellar spiral arms (white) on deprojected \sstar map.
The x and y axes are offset from the galactic centre in arcsec.
The companion galaxy M51b is located at the bottom right corner.
Middle: log-spiral stellar spiral arms (red) and predicted gas spiral arms 
for three corotation radii (asterisk, cross, diamond).
Right: predicted gas spiral arms (cross, same as the middle panel) on deprojected \sH map.
}
\label{fig:show_model}
\end{figure*}

 The stellar arm positions are defined by the results of logarithmic spiral fitting 
performed to define the arm regions in \S \ref{sec:armdef}.
 These log-spiral arms are presented in the left panel of Figure \ref{fig:show_model}, 
together with the deprojected \sstar map. 
 Based on this stellar arm positions and three $R_{\rm CR}$ estimates
($125''$, $168''$, and $300''$), 
predicted positions of gas arms are presented in the middle panel.
 The maximum radial extent of gas arms are set to be
the smaller one of $150''$ and $0.9R_{\rm CR}$.
 We find that arm2 is consistent with the galactic shock wave 
with $R_{\rm CR}=168''$, while arm1 cannot be explained by 
any of the three cases (Figure \ref{fig:show_model}, right).

\subsection{Offsets between gas and stellar spiral arms}\label{sec:ofs}
\begin{figure*}
\includegraphics[width=.475\linewidth]{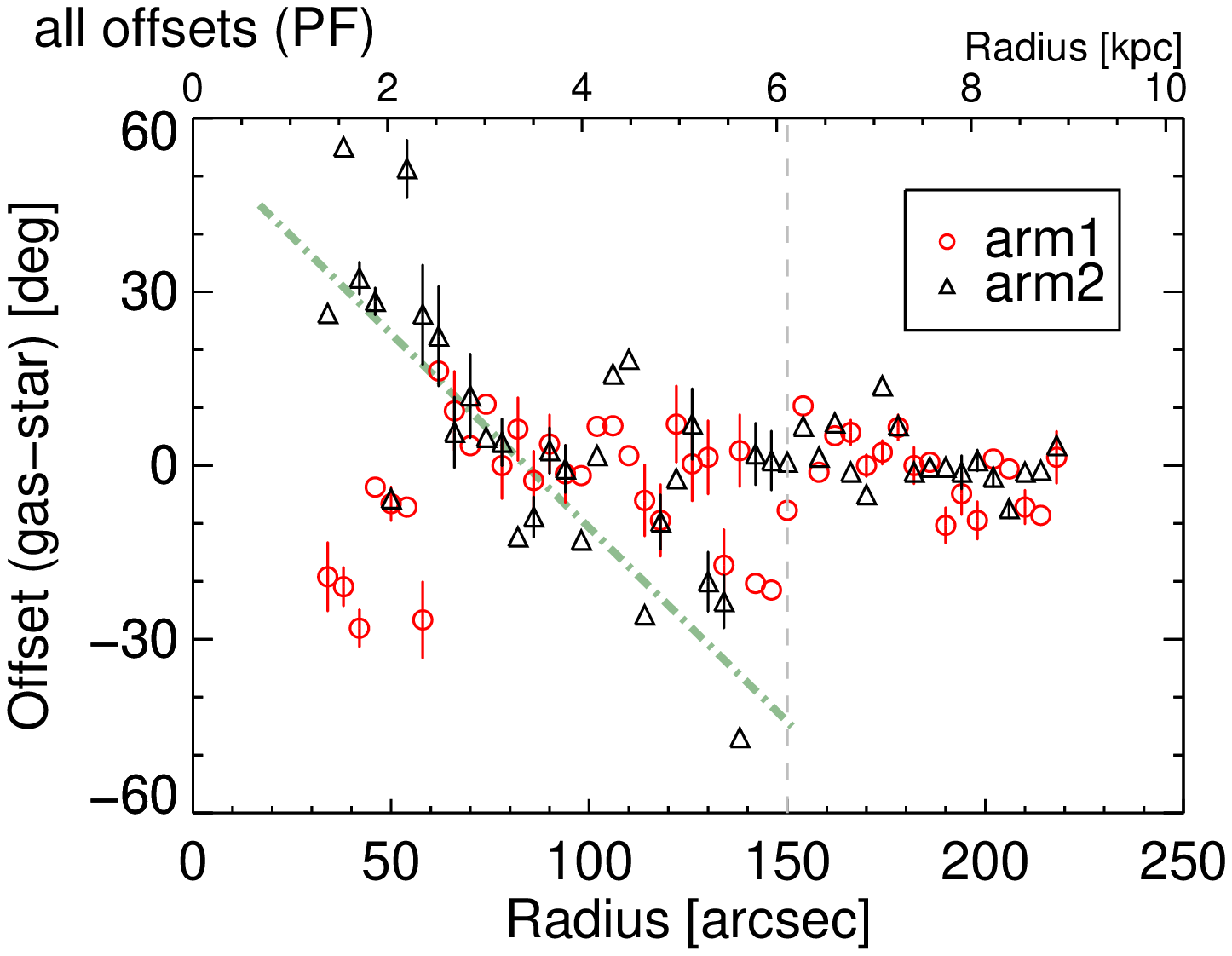}
\includegraphics[width=.475\linewidth]{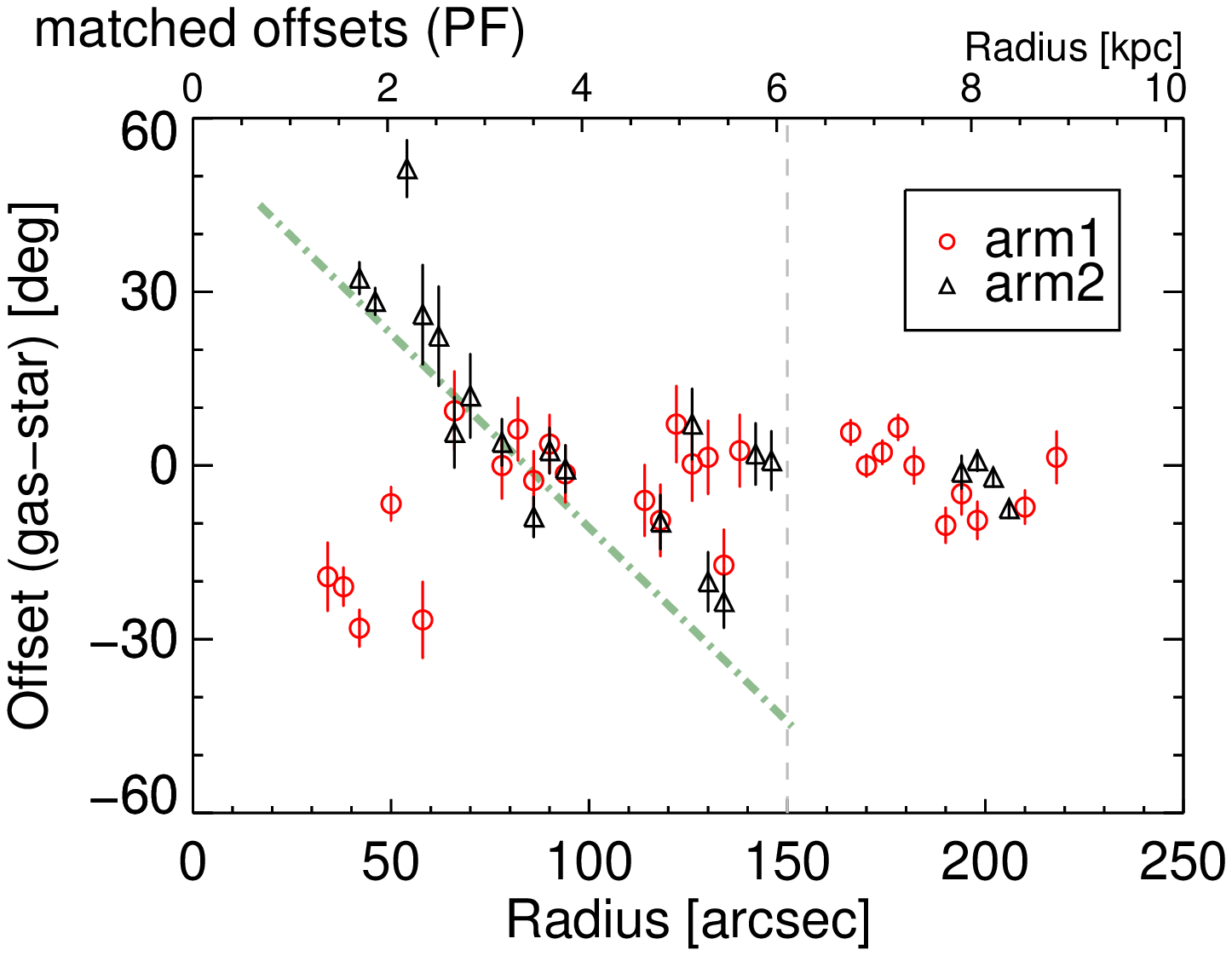}
\includegraphics[width=.475\linewidth]{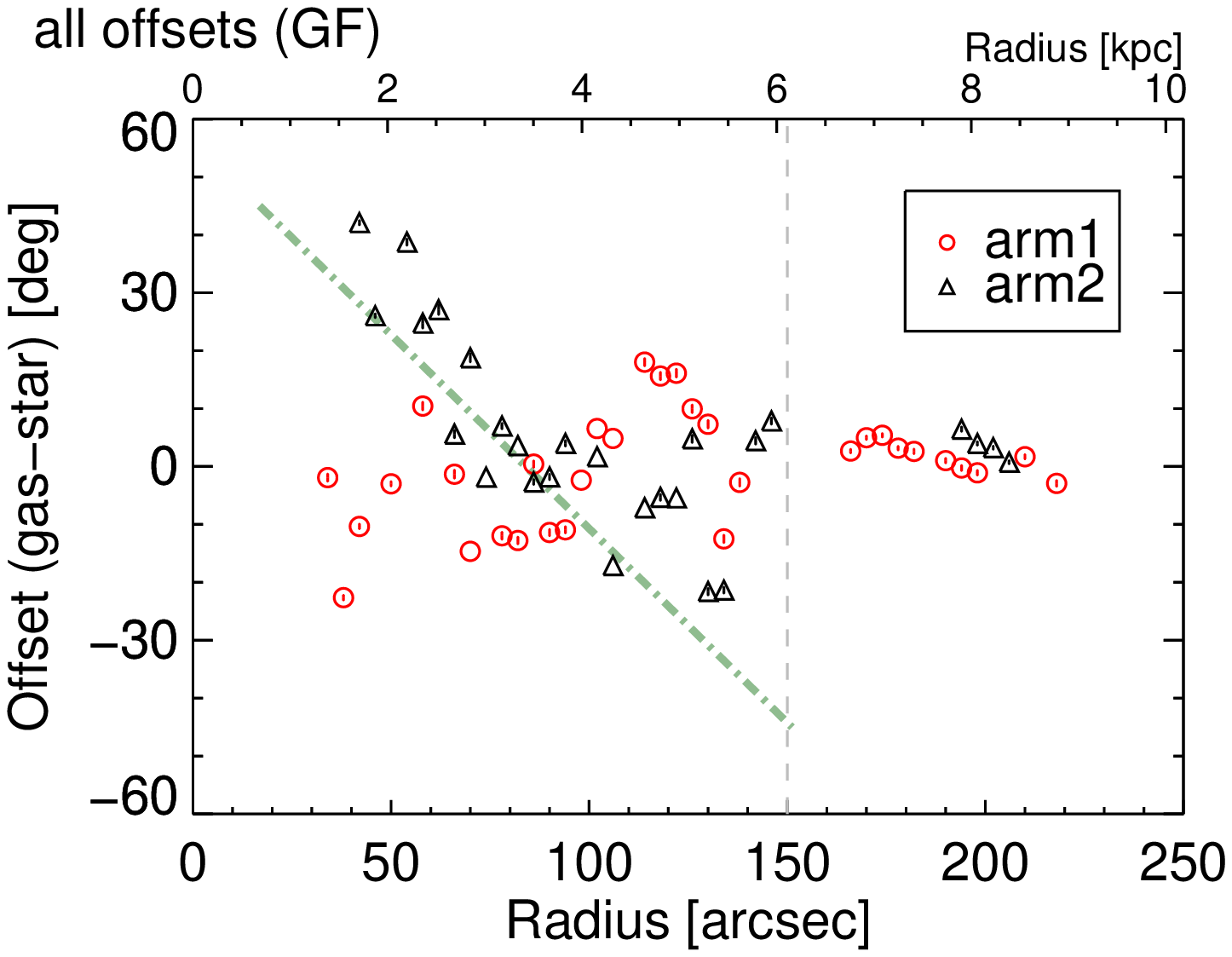}
\includegraphics[width=.475\linewidth]{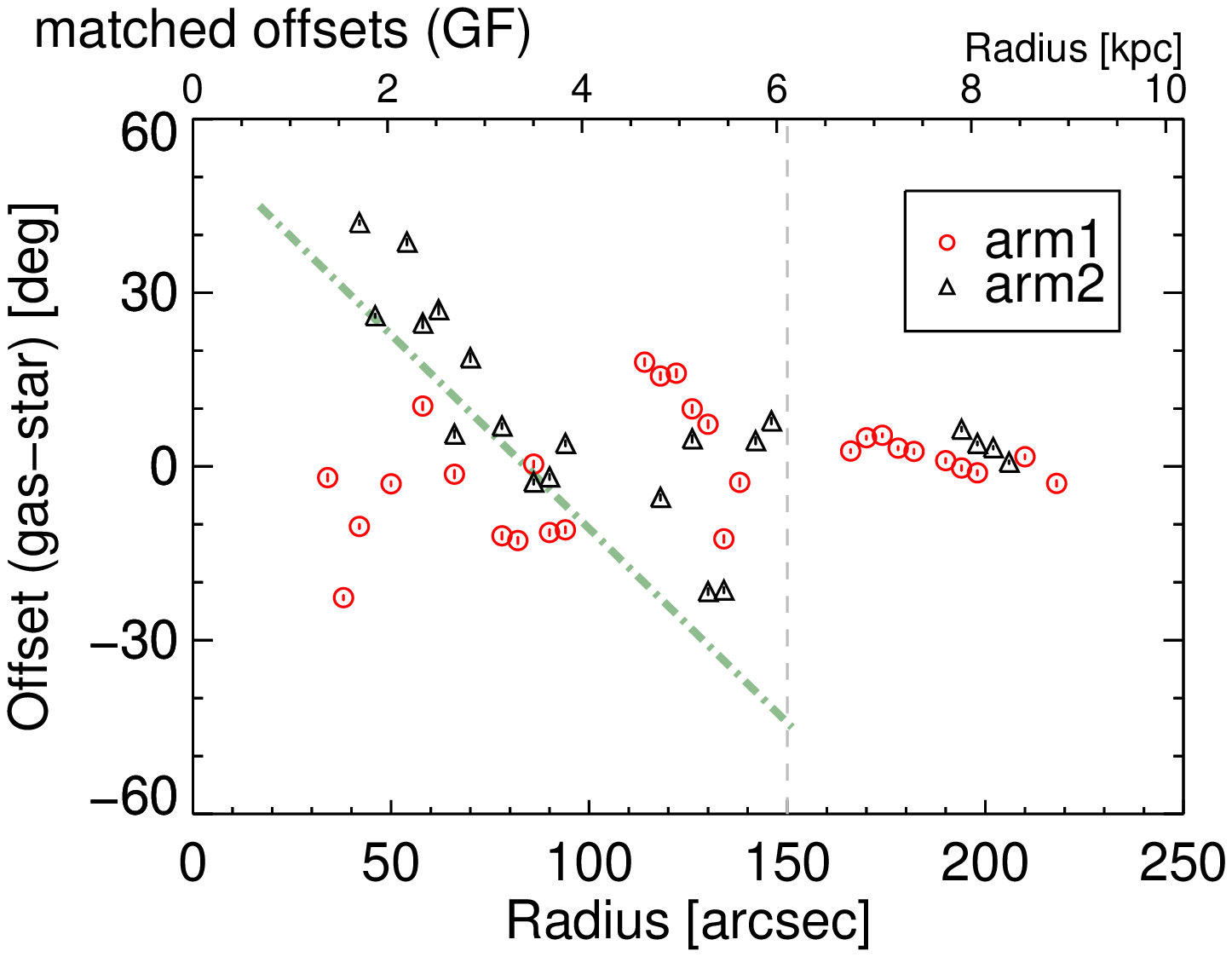}
\caption{
Offset v.s.\ radius for four different conditions. 
Top: offset from the peak-finding (PF) method. Bottom: offset from the gaussian-fit (GF) method.
Left: all offsets from each method. Right: offsets between matched peaks only.
Error bars 
\textcolor{black}{
represent random uncertainties estimated in \S \ref{sec:err}, and
}
are available only for matched offsets.
\textcolor{black}{
%Light green dot-dashed lines indicate a simplified predicted relation 
%(with $R_{\rm CR}=168''$, see \S \ref{sec:gasloc}) from galactic shock models by \citet{Baba15}.
A predicted relation from galactic shock models by \citet{Baba15} is presented as 
light green dot-dashed lines (with $R_{\rm CR}=168''$, see \S \ref{sec:gasloc}).
}
}
\label{fig:ofs_r}
\end{figure*}

 In Figure \ref{fig:ofs_r}, we plot offsets against radius
for both peak-finding (top) and gauss-fit (bottom) methods.
 Right panels are for offsets between the matched peaks only.
 Red circles and black triangles indicate arm1 and arm2, respectively.
 As the azimuthal angle is defined to increase with the same direction of galactic rotation, 
a negative offset corresponds to a gas peak at the leading 
(upstream if inside the corotation) side of a stellar peak.
 Vertical error bars are uncertainties in measured offsets estimated based on 
S/N and arm widths, which are explained in \S \ref{sec:err}.
 Since the arm widths are measured in the gaussian fitting (\S \ref{sec:gfit}), 
these error bars are 
available only for offsets between matched peaks.
 Vertical dashed line at $r=150''$ indicates the border between inner and outer arms.
\textcolor{black}{
 Light green dot-dashed lines represent the model prediction from \citet{Baba15} 
as described in \S \ref{sec:gasloc} with $R_{\rm CR}=168''$.
}
 From this plot, it is clear that the offset distribution is different for 
arm1 and arm2 as well as for the inner and outer regions.
 We find that this trend does not change depending on the methods.
 Most notable correlation is that 
offsets decrease with radius for inner arm2, 
which is consistent with the galactic shock wave.

\subsubsection{Uncertainty in offset measurements}\label{sec:err}
 In order to estimate uncertainty in measured offsets, 
we first simulate observations by the following steps: 
(1) create a map of random noise with a pixel size of $2''$ (or 82 pc), 
which is the same with \sH and \sstar maps.
(2) add a source with a 2D gaussian shape at $r=100''$ (or 4.1 kpc).
(3) convolve the map with the 2D gaussian with FWHM of $6''$, 
which is the matched PSF size used in this study.
 Free parameters of these procedures are 
the peak signal-to-noise ratio (S/N) and size of the source. 

 We then apply the same analysis described in  
\S \ref{sec:ploc} to a simulated image
and measure an azimuthal difference between identified and original peak positions.
 This process is repeated 1000 times and 
a histogram of the azimuthal difference is created.
 We fit a gaussian profile to this histogram and define $\sigma$ of the fitted gaussian 
as the uncertainty in locating peaks.

 From this simulation, we find that 
$$ \frac{\sigma}{{\rm [degree]}} \simeq \left\{
\begin{array}{cl}
\textcolor{black}{0.095} & ({\rm S/N}=3)\\
\textcolor{black}{0.075} & ({\rm S/N}=5)\\
\textcolor{black}{0.055} & ({\rm S/N}=10)
\end{array}
\right\} \times \frac{{\rm (azimuthal~source~width)}}{{\rm FWHM~[degree]}}, $$
for the peak-finding method described in \S \ref{sec:peakfind}.
 For the gaussian-fitting method (\S \ref{sec:gfit}), 
the uncertainty is about 10 times smaller.
 We should note here that this $\sigma$ is calculated for 
peak positions of gas and stars separately.
 The uncertainty in offset values is thus should be 
$\sigma_{\rm offset}=\sqrt{\sigma_{\rm H}^2+\sigma_{\rm star}^2}$.
 From S/N values derived in \S \ref{sec:stellararm} and \ref{sec:gasarm}, 
we assume ${\rm S/N}=5$ at all radii for both \sH and \sstar maps.
 Arm widths measured during the gaussian fitting to profile peaks (\S \ref{sec:gfit})
are adopted as the azimuthal source width.
 For offsets between matched peaks, 
error bars with half-length of $\sigma_{\rm offset}$ are added to the plot.
 This uncertainty is also used when calculating 
uncertainties of the correlation coefficients (\S \ref{sec:scc}).
 We should note that this uncertainty is likely underestimate, 
because it is based on idealized simulations and
does not include any systematic errors 
which are discussed in Appendix \ref{sec:syserr}.

\subsubsection{Cross correlation coefficient}\label{sec:scc}
 In order to test if the offset distribution in M51 is consistent with the galactic shock models, 
we calculate Spearman's correlation coefficient for offsets and radius 
of each arm, inner and outer arms separately.
 As well as the coefficient for all the offsets 
between all the peaks located by each method,
that for offsets between matched peaks are calculated.
 Since the latter has its uncertainty 
estimated from the arm width and S/N (\S \ref{sec:err}), 
variations of the correlation coefficient due to this uncertainty are also calculated.
 We should again note here that this uncertainty is likely underestimate.
 All of these values are listed in Table \ref{tab:scc} together with 
the number of offsets used for calculation.

\begin{table*}
\caption{
Spearman's correlation coefficients for offset and radius
}
\label{tab:scc}
\begin{tabular}{lcccccc}
\hline
\hline
\multicolumn{7}{c}{offsets from peak finding method}\\
\hline
\hline
radial range & \multicolumn{2}{c}{arm1} & \multicolumn{2}{c}{arm2} & \multicolumn{2}{c}{both arm} \\
 & all & matched & all & matched & all & matched \\
\hline
inner ($r=30''$--$150''$) & 
$0.06$ (30) & $0.41\pm 0.10$ (18) & 
$-0.65$ (30) & $-0.77\pm0.06$ (17) & 
$-0.30$ (60) & $-0.14\pm0.06$ (35)\\
outer ($r=151''$--$220''$) & 
$-0.52$ (17) & $-0.56\pm 0.20$ (10) & 
$-0.37$ (17) & NA (4) & 
$-0.47$ (34) & $-0.46\pm0.15$ (14)\\
all & 
$0.04$ (47) & $0.24\pm 0.08$ (28) & 
$-0.40$ (47) & $-0.69\pm 0.05$ (21) & 
$-0.17$ (94) & $-0.14\pm0.05$ (49)\\
\hline
\hline
\multicolumn{7}{c}{offsets from gaussian fit method}\\
\hline
\hline
radial range & \multicolumn{2}{c}{arm1} & \multicolumn{2}{c}{arm2} & \multicolumn{2}{c}{both arm} \\
 & all & matched & all & matched & all & matched \\
\hline
inner ($r=30''$--$150''$) & 
$0.40$ (22) & $0.30\pm 0.04$ (18) & 
$-0.67$ (23) & $-0.75\pm0.02$ (17) & 
$-0.15$ (45) & $-0.19\pm0.01$ (35)\\
outer ($r=151''$--$220''$) & 
$-0.84$ (10) & $-0.79\pm 0.01$ (10) & 
NA (4) & NA (4) & 
$-0.50$ (14) & $-0.51\pm0.04$ (14)\\
all & 
$0.35$ (32) & $0.30\pm 0.02$ (28) & 
$-0.51$ (27) & $-0.63\pm 0.02$ (21) & 
$-0.04$ (59) & $-0.10\pm0.01$ (49)\\
\hline
\multicolumn{7}{l}{Note: numbers in parentheses are the number of data used to calculate the coefficient.}
\end{tabular}
\end{table*}

 As already seen in Figure \ref{fig:ofs_r}, 
dependence on method is not significant.
 The largest difference is for inner arm1, 
where the coefficient for all offsets from peak finding is $\simeq 0$, 
while that for all offsets from gauss fit and for matched offsets is $\simeq 0.3$--$0.4$.
 Other arm segments, i.e.\ inner arm2 and outer arms, show negative coefficients, 
which is consistent with the galactic shock models.
 Among them, result for inner arm2 is most robust, 
as the number of measurements is largest and difference between the methods is small.
 This distribution may indicate an existence of corotation at $r> 150''$, 
as already shown in \S \ref{sec:gasarm}.
 Relationships for outer arms
are more uncertain due to offset values close to zero, 
the small number of data points, and smaller S/N.
 Especially, outer arm2 has only four offsets from matched peaks, 
which results in the failure of calculating the correlation coefficient and its variation.

\subsubsection{Pitch angles}
 By fitting the logarithmic spiral function to stellar and gas peaks, 
we measure pitch angles
for arm1 and arm2 separately as well as 
for the inner ($r=30''$--$150''$) and the outer ($r=151''$--$220''$) disk.
 The fit results are 
listed in Table \ref{tb:pitch} and
presented as blue dot(s)-dashed lines in Figure \ref{fig:peaks}.
 From this figure, it is clear that spiral arms in M51 are relatively well 
expressed by the log spiral functions within the radial ranges we used.

 Given that the galactic shock models predict gas arms move from downstream to upstream 
of stellar arms with increasing radius, a pitch angle for the gas arm should be 
smaller than that for the stellar arm.
 For M51, we find this difference in pitch angle except for inner arm1.
 This is consistent with the radial dependence of offsets presented in 
Figure \ref{fig:ofs_r} and Table \ref{tab:scc}.

\section{Discussion}
\subsection{Comparison with previous studies of M51}
 \citet{Schi13} presented the $^{12}$CO(1--0) emission data at 
$\simeq 1''$ resolution for the central $r\la 150''$ region of M51.
 Offsets between the CO and other data were calculated 
using the cross-correlation method.
 Among them, best tracers for the stellar mass are
the HST/ACS $H$-band image and 
the stellar component in Spitzer/IRAC $3.6\mu$m map from \citet{Meidt12}. 
 From their Figure 11, offsets between CO (i.e.\ molecular gas) and stars 
fall within $\pm 10^\circ$ and 
no radial dependence is seen at $r\simeq 10''$--$110''$.
 Meanwhile, we find larger offset values ($\pm 50^\circ$).
 In addition, their radial dependences are opposite for arm1 and arm2 at 
$r=30''$--$150''$.
 This discrepancy is at least partially due to the fact that 
\citet{Schi13} did not separate two arms.
 We also calculate correlation coefficients without separating two arms 
(``both arm'' column in Table \ref{tab:scc}), 
and find values between $-0.30$ and $-0.14$ for the inner region.
 \citet{Kend11} estimated the shock location in nearby spiral galaxies 
by using Spitzer/IRAC and optical images by the SINGS survey \citep{SINGS}.
 Though the scatter is relatively large, their result for M51 indicates that 
the shock is generally at the leading side of the stellar arms at $r\simeq 120''$--$250''$. %, 
 Their results are consistent with our results for inner arm2 
despite the small overlap in radius ($r=120''$--$150''$), but not for outer arms, 
where we find that offsets fluctuate around zero.
 Their pitch angle ($14^\circ$) is also smaller than our estimates 
(see Table \ref{tb:pitch}).
 While not fully understood, one reason for these inconsistencies 
could be how they derive gas and stellar peak positions.
 For gas peaks, Spitzer 8$\mu$m map was used as a tracer, 
which includes PAH features that become brighter due to emission from young stars.
 For stellar peaks, Spitzer 3.6$\mu$m and 4.5$\mu$m maps were used as a tracer.
 The authors found small-scale structures not associated with PAH features, 
and attributed them to direct emission from young stars.
 Thus, their gas and stellar peak positions could be affected by young stars 
but in a different way.

 On the other hand, information on gas kinematics has been used to 
explore the nature of spiral arms in M51.
%% CO(2-1) w/ IRAM30m, r<150" are considered
%% full map size ~4'x5' (Gar93a, Fig. 1), two arm not separated
%% no sharpening of azimuthal profile at the entrance of arms
 \citet{Gar93b} used the $^{12}$CO(2--1) emission data at 
$\sim 12''$ resolution 
for $r \la 150''$
to compare with orbit crowding model calculations 
and found that the observed arm/interarm ratio of CO brightness %% is this ``kinematic'' information??
is consistent with the model without galactic shock.
%% Kuno97: CO(1-0) w/ NRO45m, r=40-100" to derive orbit
%% difference between two arms not discussed
%% Miya14: orbits at r=40-110" and 110-140"
 \citet{Kuno97} investigated velocity distributions and orbits of gas 
from the $^{12}$CO(1--0) emission data at $\sim 15''$ resolution
for $r = 40''$--$100''$.
 Since a difference in crossing time of arm and interarm regions 
was in accordance with the arm/interarm ratio of gas density
and the velocity change across the spiral arms 
was gradual, they concluded that 
the spiral arms in M51 can be explained without galactic shock.
 Contrary to these single-dish observations, 
\citet{Aal99} claimed that interferometric CO data at higher ($\sim 2''$--$3''$) resolution 
support the presence of shocks in M51.
%% Vog93: FP Ha, 2" and 25 km/s resolution
%%  $R_{\rm CR} \simeq 130''$ for both arms from Fabry-Perot H$\alpha$ observations
%% Aal99: OVRO CO at 2-3" & ~8 km/s resolution, steep velocity gradient consistent with galactic shock
%% Hen03: BIMA CO & HST Pa-a, two arms are not symmetric, superposition of m=2&3 modes?
%% Shet07: CO from BIMA SONG+alpha, 4-6" resolution; Ha from Vog93
%%  different streaming motions due to superposition of multiple m components? 
%%  arm1 more consistent with steady-state spiral??
%%  calculated mass flux indicates arms in M51 is not steady spiral in a flat disk
 Regarding the difference between the arms, 
\citet{Vog93} suggested two spiral arms are both driven by density waves from 
H$\alpha$ residual velocities at minor and major axis. 
 Meanwhile, \citet{Shet07} presented that profiles of streaming motions for the two arms are different
based on CO and H$\alpha$ velocity fields.
 By calculating mass flow rates, they also claimed that the spiral structure in M51
is not consistent with a steady spiral model in a flat disk.
 In summary, it is still an open question whether 
quasi-steady spiral arms as well as the galactic shock waves exist in M51 or not.

\subsection{Nature of spiral arms in M51}\label{sec:lifetime}
 In this subsection, we discuss the nature of spiral arms in M51, 
taking into account offsets between \sH and \sstar presented in this paper 
as well as those between CO and H$\alpha$ studied by \citet{Egu09} and \citet{Lou13}.
 Since we find that inner and outer arms as well as arm1 and arm2 may have a different nature, 
we discuss them separately.
 Note that CO-H$\alpha$ offsets are measured at $r\simeq 40''$--$130''$, 
only for the inner region.

 First, we briefly summarize the result for the outer spiral arms.
 Their nature is not well constrained because 
(i) the number of successful gas-star offset measurements is small 
and (ii) CO-H$\alpha$ offsets are not measured.
 Despite its uncertainty, the gas-star offset for the outer arm1 
shows a relatively strong negative correlation with radius.
 This does not exclude the possibility that the outer arm1 is consistent with the galactic shock wave 
with $R_{\rm CR} \ga 300''$.
 The nature of the outer arm2 is even more uncertain, 
but the offset distribution appears similar to that of outer arm1.

%% \citet{Lou13} updated/extended this study ... radial trend not clear from their Figure 6?
 For the inner arm1, 
\citet{Egu09} measured CO-H$\alpha$ offsets and 
derived $t_{\rm SF}=7.1 \pm 0.5$ Myr, 
while $t_{\rm SF}$ is a time needed for CO clouds to form HII regions 
and is called star formation time-scale.
 Assumptions in their analysis are a constant pattern speed (i.e.\ the arm is a density wave), 
circular orbits, and constant $t_{\rm SF}$.
 Their result indicates that arm1 is stable (or rigidly rotating) at least for 
$> t_{\rm SF} \sim 10$ Myr.
 On the other hand, we find that the gas-star offset is 
unlikely to be consistent with the galactic shock models.
 This result suggests that the spiral structure is not quasi-stationary
and/or the gas self-gravity is too strong to form the standing shock waves.
 The former implies that the inner arm1 is slowly-winding.
%% Meidt08: do not handle two arms separately
%% Dob10,Stru11: no significant difference between two arms
 Its lifetime should be long enough ($\ga 10$ Myr) to have CO-H$\alpha$ offsets
consistent with the density wave and 
short enough ($\la 500$ Myr) to have gas-star offsets inconsistent with the galactic shock.
 This is in agreement with a picture proposed by numerical simulations for an interacting system 
\citep{Dob10,Stru11,Pett16}, in which spiral arms change their shape at a time-scale of a few 100 Myr.
 The latter, i.e.\ the effect of gas self-gravity, is discussed in Appendix \ref{sec:alpha}, 
but is not conclusive from the current data set.

 Interpretation of the results for the inner arm2 is not straightforward.
 While we find that the gas-star offset is consistent with the galactic shock wave, 
the CO-H$\alpha$ offset measured by \citet{Egu09} is not consistent with the density wave.
 If we follow the discussion on the inner arm1, the former suggests that 
the inner arm2 is long-lived while the latter suggests that it is short-lived.
 One possibility is that the assumptions of constant pattern speed, 
circular orbits, and constant $t_{\rm SF}$ adopted in \citet{Egu09} are not valid.
 Naively speaking, our interpretation of the inner arm1 to be slowly-winding spiral arm means that 
the pattern speed decreases with radius.
 If this is the case, a decreasing trend of CO-H$\alpha$ offsets with radius 
will be weaker.
 Furthermore, if $t_{\rm SF}$ increases with radius 
(due to lower gas density and/or metallicity at larger radii, for example), 
it will help CO-H$\alpha$ offsets become larger at larger radii.
 However, these two possibilities do not explain the difference between arm1 and arm2.
 Based on the method of \citet{Kuno97}, 
\citet{Miya14} investigated gas dynamics at $r=40''$--$110''$
and $r=110''$--$140''$.
 From their velocity vectors and profiles at $r=110''$--$140''$, 
velocity component perpendicular to the arm 
becomes larger after the passage of arm2, 
while it becomes smaller after the passage of arm1.
 Although such a difference is not seen at $r=40''$--$110''$, 
this behaviour can explain the large CO-H$\alpha$ offsets for arm2 at larger radii.
 For further discussion on the nature of inner arm2, 
more careful investigation of gas orbits is necessary, 
which is beyond the scope of this paper.

\section{Summary}
 From theoretical calculations of gas response to a quasi-stationary spiral potential, 
interstellar gas should experience a galactic shock 
around spiral arms, and offsets between the shocks and potential minima are 
expected to depend on galactocentric radius. %and the strength of self gravity of gas.
 On the other hand, no such systematic offsets are found in 
simulations of dynamic spiral structures.

 For the grand-design spiral galaxy M51, 
we have measured these offsets using high-resolution maps of
stellar mass density and hydrogen mass density.
 The former was created by fitting SED models to optical and near-infrared images, 
and thus is most reliable as uncertainties due to extinction, 
the contamination of young stars, and interstellar features are small.
 The latter was created by combining CO and HI images, 
which are the tracer of molecular and atomic hydrogen, respectively.
 The final resolution of images used in this study is $6''$, 
which corresponds to 240 pc at the adopted distance of M51.

 The offset is defined as an azimuthal difference of peak positions 
of gas and stellar mass profiles and is measured separately for two spiral arms 
at every $4''$ in radius.
 The uncertainty of offsets is estimated from the data S/N and arm width.
 We separate the inner and outer regions and 
investigate radial dependence of the offsets.
 For the inner region, arm1 and arm2 show a different dependence, 
only the latter that extends to the companion galaxy being consistent with the galactic shock wave.
 For the outer region, the stellar arms are weaker 
and offset values are closer to zero.
 The number of offsets is also smaller, so that the result is not conclusive.
 Nevertheless, the offset distributions do not exclude the possibility 
that outer arms are consistent with the galactic shock wave.
 This study highlights an importance of separating two arms 
as well as inner and outer regions for studying the nature of the spiral structure in M51.

 Furthermore, we attempt to constrain the lifetime of inner arms, 
by combining implications from gas-star offsets and 
CO-H$\alpha$ offsets \citep{Egu09,Lou13}.
 For the inner arm1, CO-H$\alpha$ offsets suggest that 
its lifetime should be longer than 
the star formation time-scale which is $\sim 10$ Myr.
 From the gas-star offsets measured in this paper, 
the inner arm1 is unlikely to be
consistent with the galactic shock wave.
 This result suggests
that its lifetime should be shorter than $\sim 500$ Myr, 
which is a time-scale to form the galactic shock, 
and/or the gas self-gravity is too strong to form the galactic shock.
 The former possibility
is consistent with numerical simulations of interacting systems, 
in which spiral arms are slowly-winding
at a time-scale of a few 100 Myr.
 Meanwhile, implications from CO-H$\alpha$ and gas-star offsets 
for the inner arm2 appear to be contradictory.
 We discuss several possibilities to cause this inconsistency.

\section*{Acknowledgments}
 Authors are grateful to careful reading and a number of comments by a referee, 
which have significantly improve 
the manuscript of this paper.
\textcolor{black}{
 This work was in part carried out on the common use data analysis computer system 
at the Astronomy Data Center, ADC, of the National Astronomical Observatory of Japan.
}
 JB was supported by HPCI Strategic Program Field 5 `The origin of 
matter and the universe'.
 JK is supported by the NSF through grant AST-1211680 for this research. 
 JK also acknowledges the support from NASA through grant NNX14AF74G.

%%% references
\bibliographystyle{mnras}
\bibliography{papers}

\appendix

\section{Systematic uncertainties}\label{sec:syserr}
\subsection{Effect of conversion factor variation to gas mass}\label{sec:Xco}
 In this paper, we use a constant conversion factor 
to derive the H$_2$ gas surface density from the CO integrated intensity.
 However, it has not been confirmed to be constant.
 In particular, its dependence on gas-phase metallicity has been actively discussed 
\citep[e.g.][]{Ari96,Bola13b}, 
as it would directly affect the ratio of CO to H$_2$.
 Since the metallicity often decreases with radius, 
the conversion factor is suggested to increase with radius. 
 However, no strong radial dependence is found by \citet{DM13} 
at $r \la 3$ kpc for three nearby galaxies.
 On much smaller scales, the conversion factor is likely variable within a molecular cloud, 
since CO is more easily destroyed by UV radiation when H$_2$ is still self-shielded.
% Nevertheless, an azimuthal variation of the conversion factor has not been investigated (really?).
 \citet{Schi10} performed LVG analysis on 17 locations
(120 or 180 pc in diameter) in a spiral arm of M51 
and derived the conversion factor for each of them.
 Most of the values are consistent with the Galactic value within a factor of two.
 There might be a tendency that downstream clouds have a slightly higher conversion factor, 
perhaps due to stronger UV radiation from young stars, 
if any.
 If such a systematic variation of the conversion factor exists, 
CO intensity peaks would shift upstream from the true gas density peaks.
 Since they are systematic, these shifts cannot erase the radial trend of offsets for the galactic shock.
 Again, this trend is within the factor of two variation and not confirmed yet.
 For a metal-rich galaxy such as M51, the conversion factor is 
estimated to be constant within a factor of few except the central region 
\citep[e.g.][]{Sand13}.

%% Bola13b for review
%% works claiming dependence: 
%% Wil95 (CO(1-0) for 5 galaxies (interferemetric and single dish data co-exist?), Mvir, M33 divided into 3 regions due to metallicity gradient)
%% Ari96 (CO(1-0) for 7 galaxies + MW (interferemetric and single dish data co-exist), some galaxies divided into multiple regions, radial gradient of Xco for MW -- applicable to other galaxies?)
%% Bose02 (CO(1-0) w/ KP12m), 
%% Schr12 (CO(2-1) w/ IRAM30m, constant 2-1/1-0 ratio and SFE, galaxy-integrated)

%% works against dependence: Bli07, Bola08

%% on neutral position?: 
%% DM13 (CANON, from Mvir, no strong radial dependence within a galaxy)
%% Sand13 (CO(2-1) w/ IRAM30m & Herschel, Xco mostly flat and drop at center?, only weak correlation with metallicity)

%% drop in galactic centers: Sand13 and references in their S.5.1 

%% for M51: 
%% GB93 (Xco=0.8 from 12CO(1-0) and 13CO(1-0) w/ LVG)
%% NK95 (Xco=0.9 from Av and CO(1-0))
%% higher values from Mvir (RK90; Adl92)
%% Schi10 (LVG with 12CO(1-0)&(2-1), 13CO(1-0), Xco ~ Tkin^(-1) n(H2)^0.5?)

%% for high-z: Gen12

\subsection{Effect of attenuation to stellar mass}\label{sec:Av}
 The \sstar map used in this study is created by 
fitting SED models to optical and NIR images.
 The methodology is the same as \citet{MenC12}, 
but the angular resolution is higher than their work.
 While the stellar mass is the most robust parameter among the fitting results, 
it may be subject to the age-extinction degeneracy.
 For example, a redder SED may come from 
highly extincted star light and/or older stars.
 The former leads to an underestimation of stellar mass, 
while the latter leads to an overestimation.
 \citet{MenC12} compared their $A_V$ values with \citet{Muno09}, 
who used the IR-to-UV ratio to derive the amount of extinction.
 While the most of these values are consistent within their uncertainties, 
$A_V$ values from the SED fitting are on average $\sim 0.2$ mag lower 
than those from the IR/UV ratio.
 If UV and FIR brightness were included in the SED fitting, 
we would be able to better constrain the stellar mass distribution.
 However, %FIR images at $4''$ resolution are not available.
the current highest resolution of FIR images is 
$\sim 20''$ achieved by Herschel, 
which is significantly larger than the resolution of \sstar and \sH 
maps ($6''$) used in this study.

 By examining the \sstar and \sH maps, 
we find a dip in \sstar where \sH peaks in several locations.
 So it is more likely that the stellar mass is underestimated
when the gas density is high and thus the extinction is large.
 This effect should create offsets even if 
peaks of true stellar mass and gas mass coincide, 
but such artificial offsets should appear in random direction.
 If true gas peaks move from downstream to upstream with increasing radius, 
as predicted by the galactic shock models, 
the attenuation effect would shift stellar peaks to the opposite side of gas peaks.
 This results in observed offsets to be larger than true offsets, 
but the radial trend does not change.

 In addition, we locate spiral arms by fitting a gaussian profile 
to the azimuthal profiles in \S \ref{sec:gfit}.
 As the width of fitted gaussian for stellar arms is wider than that for gas arms, 
we deduce that the derived stellar arm positions are minimally affected by 
the attenuation due to gas arms.
 If stellar peaks were artificially created by over correcting the extinction, 
widths of gas and stellar arms would be similar.
%% matched/all = (33+30)/(47+47) = 0.67
% At least for the inner arms, we confirm in \S \ref{sec:p-gfit} that the radial trend of offsets 
%from the gaussian fitting technique is consistent with that from the peak locating technique.
 We confirm in \S \ref{sec:ofs} that the radial trend of offsets 
from the peak-finding method and gauss-fit method is consistent with each other.
 We thus conclude that the attenuation effect does not significantly change 
the trend of gas-star offset against radius presented in this paper.

\section{Dependence on adopted parameters}\label{sec:params}
\subsection{Radial ranges}
 Here we test how adopted radial ranges affect the result. 
 The first test is to treat inner and outer arms as a single arm.
 The cross correlation coefficients in this case are listed in 
``radial range = all'' rows in Table \ref{tab:scc}.
 The coefficients for arm1 are positive, and thus a hint of negative correlation 
for the outer arm disappears.
 The coefficients for arm2 do not differ, as both inner and outer arms show negative correlations 
as well as the number of outer arm measurements is small.

 The next test is to exclude the innermost region ($r\lesssim 55''$). 
 In this region, we find stellar peaks deviate from the logarithmic spiral (Figure \ref{fig:peaks}).
 It is likely due to that azimuthal \sstar profiles are rather flat, 
resulting in stellar peak positions being more uncertain.
 In addition, several studies suggested that 
the innermost spiral structure is in a different dynamical condition \citep{Meidt13,Colo14b}.
 When this region is excluded, the cross-correlation coefficient for inner arm1 
becomes slightly negative ($\sim -0.1$) for the peak-finding method and 
stays positive for the gaussian-fitting method.
% This indicates that the result for inner arm1 is not robust?
 The effect of excluding the innermost region is not significant, 
but it highlights again the difference between the methods for inner arm1.
 Results for inner arm2 do not significantly change, 
%if the innermost region is excluded.
i.e.\ all coefficients stay negative and are consistent with the galactic shock.

%\textcolor{black}{
% The effect of adopted radial ranges can be also seen in 
%``radial range = all'' rows in Table \ref{tab:scc}.
% When inner and outer arms are treated as a single arm, 
%arm1 shows positive correlation coefficients for the both methods.
% This indicates importance of setting appropriate radial ranges?
%}

\subsection{Arm regions}
 In our analysis, \sstar and \sH peak locations are restricted to 
be within the defined arm regions.
 Here we test how the width of outer arm regions affect the result.
 When we expand the outer arm region width from $30^\circ$ to $45^\circ$, 
the number of gauss-fit peaks for outer arm2 increases from four to eight.
 The cross-correlation coefficient for the gauss-fit method becomes positive ($0.55$), 
while that for the peak-finding method stays negative ($-0.22$).
 Nevertheless, the number of measurements is still small to derive a robust conclusion.
 Results for outer arm1 are almost the same with the original arm definition.

\section{Effect of gas self-gravity}\label{sec:alpha}
 As already mentioned in \S \ref{sec:intro}, 
several theoretical studies based on 1D calculations suggested that the self-gravity of gas 
shifts the galactic shock location downstream \citep{Lub86,KO02,LeeWK14}.
 On the other hand, 2D numerical simulations by \citet{Baba15} showed that 
shock locations with and without self-gravity almost coincide.
 Several possible reasons for this discrepancy are: 
1) difference in models (e.g.\ 1D vs 2D),
2) shift due to the self-gravity is small. 
 In Figure 6 of \citet{LeeWK14}, 
%the shift is $\sim 200$ pc perpendicular to a spiral arm at $r=2$ kpc, 
%which corresponds to $\sim 2^\circ$ in azimuth.
the difference in peak positions for the strong and weak self-gravity cases 
is $\sim 200$ pc perpendicular to a spiral arm.
 Given $r=2$ kpc and $i_{\rm pitch}=21^\circ$ in the model, 
this shift corresponds to $\sim 2^\circ$ in azimuthal direction.
 Meanwhile, the uncertainty in offset from the peak-finding method is 
typically $\sim 5^\circ$.
3) feedback effect: the model with self-gravity in \citet{Baba15} includes stellar feedback.
This affects gas density distribution especially at the downstream side 
and might be able to shift gas peaks upstream.

 From observational point of view, 
it is still difficult to constrain the self-gravity strength.
 \citet{LeeWK14} defined it as 
% In \citet{LeeWK14}, the self-gravity strength is defined as 
$$\alpha = \frac{2\pi mG\Sigma_0}{r \kappa^2\sin (i_{\rm pitch})},$$
where $m$ is the number of spiral arms, 
$\Sigma_0$ is an average gas density, %$i_{\rm pitch}$ is the pitch angle, 
and $\kappa$ is the epicyclic frequency.
 Therefore, 
$\alpha$ depends on $X_{\rm CO}$, $\kappa$, $i_{\rm pitch}$, and distance.
 As already discussed in Appendix \ref{sec:Xco}, 
the $X_{\rm CO}$ uncertainty is a factor of few.
 We estimate the $\alpha$ uncertainty due to uncertainties in $\kappa$ and $i_{\rm pitch}$ 
to be $<10$\% for each.
 The uncertainty in distance is $\sim 15$\%, according to NED distances.
 As a result, $\alpha$ is uncertain at least by a factor of two.

\begin{figure}
\includegraphics[width=\linewidth]{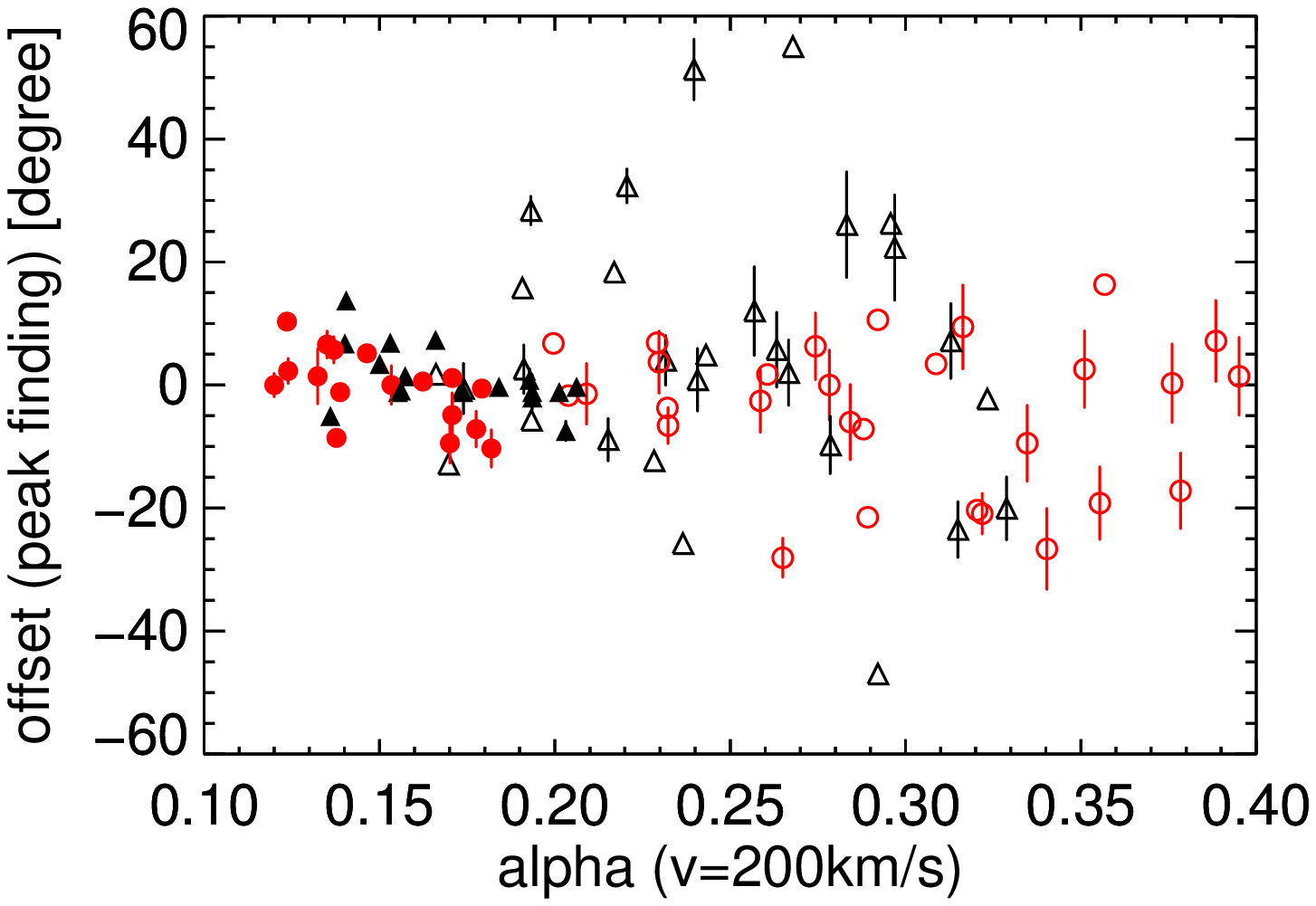}
\caption{
Offset from the peak-finding method against $\alpha$ assuming a flat rotation. 
Open and filled symbols are for inner and outer arms, respectively.
}
\label{fig:alofs}
\end{figure}

 For M51, we calculate $\alpha$ using the constant $X_{\rm CO}$, 
$\kappa$ assuming a constant rotational velocity $V_{\rm rot}=200$ km/s, 
stellar $i_{\rm pitch}$ from Table \ref{tb:pitch}, 
and distance from Table \ref{tb:m51}.
 The $\alpha$ ranges for inner and outer arms are 0.2--0.4 and 0.1--0.2, 
respectively.
% Given this small dynamic range, the uncertainty of a factor of $>2$ is large ...
 The offset from the peak-finding method is plotted against $\alpha$ in Figure \ref{fig:alofs}.
 We do not add error bars for $\alpha$, since it is uncertain as well as 
too large compared to its dynamic range.
% While the results of \citet{LeeWK14} suggest that they should positively correlate, 
%we do not find such a correlation in M51.
 If the gas self-gravity shifts a galactic shock front downstream,
this shift should be larger at inner radii as $\alpha$ generally decreases with radius.
 Therefore, the decreasing trend of offset with radius, which is predicted by the galactic shock models, 
should not be changed due to the presence of self-gravity.
 An analogy is true for the correlation between offset and $\alpha$.
 Offset should increase with $\alpha$ if both offset and $\alpha$ decrease with radius.
 This trend is not dependent of whether the self-gravity effect is included or not.
 However, such a positive correlation is not clear in Figure \ref{fig:alofs}, 
even for inner arm2 (black open triangles), where 
the radial dependence of offsets is consistent with the galactic shock models.
 We attribute this discrepancy to the facts that 
$\alpha$ is not monotonically decreasing with radius at smaller scales
and that the uncertainty in $\alpha$ is again large compared to its small dynamic range.

 Another theoretical indication %by \citet{LeeWK14} 
is that 
if the self-gravity is too strong, no solution for the galactic shock is found.
 The threshold in \citet{LeeWK14} is $\alpha \sim 0.2$, depending on other parameters 
such as a magnetic field strength.
% Note: his calculation is only at $r=2$ kpc.
% Given the large uncertainty in $\alpha$, 
%it is not clear if the reason for the lack of positive correlation in M51 
%is because $\alpha$ exceeds the threshold.
%% The inner arm1 shows slightly larger $\alpha$ than inner arm2.
 Larger $\alpha$ values for inner arm1 than for inner arm2 
can be one reason for the lack of galactic shock in inner arm1.
 We should again note here that the difference in $\alpha$ is not large 
compared to its uncertainty.
% Furthermore, $\alpha$ values in \citet{Baba15} simulations are $\sim 0.05$--$0.1$, 
%which is smaller than the threshold, but no systematic shift is found.
 
\bsp

\label{lastpage}

\end{document}